%physmatex1 contains two .sty files, the present one and pmtformats1.sty, and
%a manual, manual_pmt1.tex

%this is pmt20.sty, version August 1998.
%by F. J. Yndur\'ain
%Universidad Aut\'onoma de Madrid,
%Canto Blanco, 28049-Madrid.
%e-mail: fjy@delta.ft.uam.es
%comments welcome.

%-----------------------------------------------------------------------------

%-----------------------------

%Macros for typesetting formulas and text for mathematics 
%and physics. To be used in math mode (some 
% work outside it). Some macros are original; others have been 
%taken from eplain.tex (Karl Berry et al.), 
%manmac.tex, the \TeXtbook (D. Knuth), or Springer-Verlag macros.

%----------------------------------------------------------------------------
%------------------------------------------------------------------------------

%SUPPLEMENTARY FONTS.
%---------------------------------------
 %text italic

%--------------------------------------------------------------------
%Fonts for authors, in references

 %caps and small caps
%----------------------------------------------------------------------------
%"blackboard" (uppercase) fonts:

   \def\bbbc{{\mathchoice {\setbox0=\hbox{$\displaystyle\rm C$}\hbox{\hbox
   to0pt{\kern0.4\wd0\vrule height0.9\ht0\hss}\box0}}
   {\setbox0=\hbox{$\textstyle\rm C$}\hbox{\hbox
   to0pt{\kern0.4\wd0\vrule height0.9\ht0\hss}\box0}}
   {\setbox0=\hbox{$\scriptstyle\rm C$}\hbox{\hbox
   to0pt{\kern0.4\wd0\vrule height0.9\ht0\hss}\box0}}
   {\setbox0=\hbox{$\scriptscriptstyle\rm C$}\hbox{\hbox
   to0pt{\kern0.4\wd0\vrule height0.9\ht0\hss}\box0}}}}

   \def\bbbg{{\mathchoice {\setbox0=\hbox{$\displaystyle\rm G$}\hbox{\hbox
   to0pt{\kern0.4\wd0\vrule height0.9\ht0\hss}\box0}}
   {\setbox0=\hbox{$\textstyle\rm G$}\hbox{\hbox
   to0pt{\kern0.4\wd0\vrule height0.9\ht0\hss}\box0}}
   {\setbox0=\hbox{$\scriptstyle\rm G$}\hbox{\hbox
   to0pt{\kern0.4\wd0\vrule height0.9\ht0\hss}\box0}}
   {\setbox0=\hbox{$\scriptscriptstyle\rm G$}\hbox{\hbox
   to0pt{\kern0.4\wd0\vrule height0.9\ht0\hss}\box0}}}}

   \def\bbbo{{\mathchoice {\setbox0=\hbox{$\displaystyle\rm O$}\hbox{\raise
   0.15\ht0\hbox to0pt{\kern0.4\wd0\vrule height0.8\ht0\hss}\box0}}
   {\setbox0=\hbox{$\textstyle\rm O$}\hbox{\raise
   0.15\ht0\hbox to0pt{\kern0.4\wd0\vrule height0.8\ht0\hss}\box0}}
   {\setbox0=\hbox{$\scriptstyle\rm O$}\hbox{\raise
   0.15\ht0\hbox to0pt{\kern0.4\wd0\vrule height0.7\ht0\hss}\box0}}
   {\setbox0=\hbox{$\scriptscriptstyle\rm O$}\hbox{\raise
   0.15\ht0\hbox to0pt{\kern0.4\wd0\vrule height0.7\ht0\hss}\box0}}}}

   \def\bbbq{{\mathchoice {\setbox0=\hbox{$\displaystyle\rm Q$}\hbox{\raise
   0.15\ht0\hbox to0pt{\kern0.4\wd0\vrule height0.8\ht0\hss}\box0}}
   {\setbox0=\hbox{$\textstyle\rm Q$}\hbox{\raise
   0.15\ht0\hbox to0pt{\kern0.4\wd0\vrule height0.8\ht0\hss}\box0}}
   {\setbox0=\hbox{$\scriptstyle\rm Q$}\hbox{\raise
   0.15\ht0\hbox to0pt{\kern0.4\wd0\vrule height0.7\ht0\hss}\box0}}
   {\setbox0=\hbox{$\scriptscriptstyle\rm Q$}\hbox{\raise
   0.15\ht0\hbox to0pt{\kern0.4\wd0\vrule height0.7\ht0\hss}\box0}}}}

   \def\bbbt{{\mathchoice {\setbox0=\hbox{$\displaystyle\rm
   T$}\hbox{\hbox to0pt{\kern0.3\wd0\vrule height0.9\ht0\hss}\box0}}
   {\setbox0=\hbox{$\textstyle\rm T$}\hbox{\hbox
   to0pt{\kern0.3\wd0\vrule height0.9\ht0\hss}\box0}}
   {\setbox0=\hbox{$\scriptstyle\rm T$}\hbox{\hbox
   to0pt{\kern0.3\wd0\vrule height0.9\ht0\hss}\box0}}
   {\setbox0=\hbox{$\scriptscriptstyle\rm T$}\hbox{\hbox
   to0pt{\kern0.3\wd0\vrule height0.9\ht0\hss}\box0}}}}

   \def\bbbs{{\mathchoice
   {\setbox0=\hbox{$\displaystyle     \rm S$}\hbox{\raise0.5\ht0\hbox
   to0pt{\kern0.35\wd0\vrule height0.45\ht0\hss}\hbox
   to0pt{\kern0.55\wd0\vrule height0.5\ht0\hss}\box0}}
   {\setbox0=\hbox{$\textstyle        \rm S$}\hbox{\raise0.5\ht0\hbox
   to0pt{\kern0.35\wd0\vrule height0.45\ht0\hss}\hbox
   to0pt{\kern0.55\wd0\vrule height0.5\ht0\hss}\box0}}
   {\setbox0=\hbox{$\scriptstyle      \rm S$}\hbox{\raise0.5\ht0\hbox
   to0pt{\kern0.35\wd0\vrule height0.45\ht0\hss}\raise0.05\ht0\hbox
   to0pt{\kern0.5\wd0\vrule height0.45\ht0\hss}\box0}}
   {\setbox0=\hbox{$\scriptscriptstyle\rm S$}\hbox{\raise0.5\ht0\hbox
   to0pt{\kern0.4\wd0\vrule height0.45\ht0\hss}\raise0.05\ht0\hbox
   to0pt{\kern0.55\wd0\vrule height0.45\ht0\hss}\box0}}}}

   \def\bbbu{{\mathchoice {\setbox0=\hbox{$\displaystyle\rm U$}\hbox{\hbox
   to0pt{\kern0.4\wd0\vrule height0.9\ht0\hss}\box0}}
   {\setbox0=\hbox{$\textstyle\rm U$}\hbox{\hbox
   to0pt{\kern0.4\wd0\vrule height0.9\ht0\hss}\box0}}
   {\setbox0=\hbox{$\scriptstyle\rm U$}\hbox{\hbox
   to0pt{\kern0.4\wd0\vrule height0.9\ht0\hss}\box0}}
   {\setbox0=\hbox{$\scriptscriptstyle\rm U$}\hbox{\hbox
   to0pt{\kern0.4\wd0\vrule height0.9\ht0\hss}\box0}}}}

   \def\bbbz{{\mathchoice {\hbox{$\textstyle Z\kern-0.4em Z$}}
   {\hbox{$\textstyle Z\kern-0.4em Z$}}
   {\hbox{$\scriptstyle Z\kern-0.3em Z$}}
   {\hbox{$\scriptscriptstyle Z\kern-0.2em Z$}}}}

%"blackboard" style number 1 (\bbbone) and letters b,c,d,e,...z with 
%control sequence \bbb<letter>:
%\bbbone,\bbbb,\bbbc,\bbbd,\bbbe,\bbbf,\bbbg,\bbbh,\bbbi,\bbbk,\bbbl,
%\bbbm,\bbbn,\bbbo,\bbbp, \bbbq,\bbbr,\bbbs,\bbbt,\bbbu,\bbbz 
%note that NOT all letters are represented: the missing ones are 
% A, J, V, W, Y

%----------------------------------------------------------
%Springer \tens, \bvec, \petit fonts.

      \font \ninebf                 = cmbx9
      \font \ninei                  = cmmi9
      \font \nineit                 = cmti9
      \font \ninerm                 = cmr9
      \font \ninesans               = cmss10 at 9pt
      \font \ninesl                 = cmsl9
      \font \ninesy                 = cmsy9
      \font \ninett                 = cmtt9
      \font \fivesans               = cmss10 at 5pt
						\font \sevensans              = cmss10 at 7pt
      \font \sixbf                  = cmbx6
      \font \sixi                   = cmmi6
      \font \sixrm                  = cmr6
						\font \sixsans                = cmss10 at 6pt
      \font \sixsy                  = cmsy6
      \font \tams                   = cmmib10
      \font \tamss                  = cmmib10 scaled 700
						\font \tensans                = cmss10
    
%---------------------------------------------------------------------------
%---------------------------------------------------------------------------
  % petit-fonts
      \skewchar\ninei='177 \skewchar\sixi='177
      \skewchar\ninesy='60 \skewchar\sixsy='60
      \hyphenchar\ninett=-1
      \def\newline{\hfil\break}%
 %--------------------------------------------------------------------------
      \catcode`@=11
      \def\folio{\ifnum\pageno<\z@
      \uppercase\expandafter{\romannumeral-\pageno}%
      \else\number\pageno \fi}
      \catcode`@=12 % at signs are no longer letters

 %---------------------------------------------------------------------------
      \newfam\sansfam
      \textfont\sansfam=\tensans\scriptfont\sansfam=\sevensans
      \scriptscriptfont\sansfam=\fivesans
      \def\sans{\fam\sansfam\tensans}
 %---------------------------------------------------------------------------
\def\tens#1{\relax\ifmmode
\mathchoice{\hbox{$\displaystyle\sans#1$}}{\hbox{$\textstyle\sans#1$}}
{\hbox{$\scriptstyle\sans#1$}}{\hbox{$\scriptscriptstyle\sans#1$}}\else
$\sans#1$\fi}

%sans-serif math mode fonts for tensors.

%------------------------------------------------------------------------
%-----------------------------------------------------------------------------

      \def\petit{\def\rm{\fam0\ninerm}%
      \textfont0=\ninerm \scriptfont0=\sixrm \scriptscriptfont0=\fiverm
       \textfont1=\ninei \scriptfont1=\sixi \scriptscriptfont1=\fivei
       \textfont2=\ninesy \scriptfont2=\sixsy \scriptscriptfont2=\fivesy
       \def\it{\fam\itfam\nineit}%
       \textfont\itfam=\nineit
       \def\sl{\fam\slfam\ninesl}%
       \textfont\slfam=\ninesl
       \def\bf{\fam\bffam\ninebf}%
       \textfont\bffam=\ninebf \scriptfont\bffam=\sixbf
       \scriptscriptfont\bffam=\fivebf
       \def\sans{\fam\sansfam\ninesans}%
       \textfont\sansfam=\ninesans \scriptfont\sansfam=\sixsans
       \scriptscriptfont\sansfam=\fivesans
       \def\tt{\fam\ttfam\ninett}%
       \textfont\ttfam=\ninett
       \normalbaselineskip=11pt
       \setbox\strutbox=\hbox{\vrule height7pt depth2pt width0pt}%
       \normalbaselines\rm

%----------------------------------------------------------------------

      \def\bvec##1{{\textfont1=\tbms\scriptfont1=\tbmss
      \textfont0=\ninebf\scriptfont0=\sixbf
      \mathchoice{\hbox{$\displaystyle##1$}}{\hbox{$\textstyle##1$}}
      {\hbox{$\scriptstyle##1$}}{\hbox{$\scriptscriptstyle##1$}}}}}

%------------------------------------------------------------------------------

\font\teneufm=eufm10
\font\seveneufm=eufm7
\font\fiveeufm=eufm5
\newfam\eufmfam
\textfont\eufmfam=\teneufm
\scriptfont\eufmfam=\seveneufm
\scriptscriptfont\eufmfam=\fiveeufm

%gothic characters (Fraktur)
%------------------------------------
%-----------------------------------

%The following characters are defined for compatibility with \AMSTeX and Springer
%font macros. That is to say, they can be used inside, or outside those 
%formats.

					\mathchardef\gammav="0100
     \mathchardef\deltav="0101
     \mathchardef\thetav="0102
     \mathchardef\lambdav="0103
     \mathchardef\xiv="0104
     \mathchardef\piv="0105
     \mathchardef\sigmav="0106
     \mathchardef\upsilonv="0107
     \mathchardef\phiv="0108
     \mathchardef\psiv="0109
     \mathchardef\omegav="010A

%"versal", or slanted upper case greek characters

					\mathchardef\gammau="0000
     \mathchardef\deltau="0001
     \mathchardef\thetau="0002
     \mathchardef\lambdau="0003
     \mathchardef\xiu="0004
     \mathchardef\piu="0005
     \mathchardef\sigmau="0006
     \mathchardef\upsilonu="0007
     \mathchardef\phiu="0008
     \mathchardef\psiu="0009
     \mathchardef\omegau="000A

%Upright upper case greek characters
%--------------------------------
%The same characters are obtained with the following  definitions:
					\mathchardef\Gammav="0100
     \mathchardef\Deltav="0101
     \mathchardef\Thetav="0102
     \mathchardef\Lambdav="0103
     \mathchardef\Xiv="0104
     \mathchardef\Piv="0105
     \mathchardef\Sigmav="0106
     \mathchardef\Upsilonv="0107
     \mathchardef\Phiv="0108
     \mathchardef\Psiv="0109
     \mathchardef\Omegav="010A

%"versal", or slanted upper case greek characters

					\mathchardef\Gammau="0000
     \mathchardef\Deltau="0001
     \mathchardef\Thetau="0002
     \mathchardef\Lambdau="0003
     \mathchardef\Xiu="0004
     \mathchardef\Piu="0005
     \mathchardef\Sigmau="0006
     \mathchardef\Upsilonu="0007
     \mathchardef\Phiu="0008
     \mathchardef\Psiu="0009
     \mathchardef\Omegau="000A

%Upright upper case greek characters
%---------------------------------------

\font\grbfivefm=cmbx5
\font\grbsevenfm=cmbx7
\font\grbtenfm=cmbx10 %for the greek bf family
\newfam\grbfam
\textfont\grbfam=\grbtenfm
\scriptfont\grbfam=\grbsevenfm
\scriptscriptfont\grbfam=\grbfivefm

\font\calbfivefm=cmbsy10 at 5pt
\font\calbsevenfm=cmbsy10 at 7pt
\font\calbtenfm=cmbsy10 %for the cal bf family
\newfam\calbfam
\textfont\calbfam=\calbtenfm
\scriptfont\calbfam=\calbsevenfm
\scriptscriptfont\calbfam=\calbfivefm

%boldface for upper case upright greek characters (\grbf)
% and upper case \cal characters (\calbf)

%-----------------------------------

      \def\bvec#1{{\textfont1=\tams\scriptfont1=\tamss
      \textfont0=\tenbf\scriptfont0=\sevenbf
      \mathchoice{\hbox{$\displaystyle#1$}}{\hbox{$\textstyle#1$}}
      {\hbox{$\scriptstyle#1$}}{\hbox{$\scriptscriptstyle#1$}}}}

%boldface for slanted latin and slanted greek characters. The 
%notation \rvec, reserved for arrow over character; see below
%------------------------------------

%-----------------------------------------------------------------------------

\def\pmbf#1{\leavevmode\setbox0=\hbox{#1}%
\kern-.02em\copy0\kern-\wd0
\kern.04em\copy0\kern-\wd0
\kern-.02em\copy0\kern-\wd0
\kern-.03em\copy0\kern-\wd0
\kern.06em\box0 }

%"poor man" boldfaces, with automatic math scaling.

%-----------------------------------------------------------------------------
%--------------------------------------------------------------------------
%TEXT MACROS
%-------------------------------------------------------------------------
						%time display with command \timestamp

						\def\monthname{%
   			\ifcase\month
      \or Jan\or Feb\or Mar\or Apr\or May\or Jun%
      \or Jul\or Aug\or Sep\or Oct\or Nov\or Dec%
   			\fi
							}%
					\def\timestring{\begingroup
   		\count0 = \time
   		\divide\count0 by 60
   		\count2 = \count0   % The hour, from zero to 23.
   		\count4 = \time
   		\multiply\count0 by 60
   		\advance\count4 by -\count0   % The minute, from zero to 59.
   		\ifnum\count4<10
     \toks1 = {0}%
   		\else
     \toks1 = {}%
   		\fi
   		\ifnum\count2<12
      \toks0 = {a.m.}%
   		\else
      \toks0 = {p.m.}%
      \advance\count2 by -12
   		\fi
   		\ifnum\count2=0
      \count2 = 12
   		\fi
   		\number\count2:\the\toks1 \number\count4 \thinspace \the\toks0
					\endgroup}%

				\newskip\abovelistskip      \abovelistskip = .5\baselineskip 
				\newskip\interitemskip      \interitemskip = 0pt
				\newskip\belowlistskip      \belowlistskip = .5\baselineskip
				\newdimen\listleftindent    \listleftindent = 0pt
				\newdimen\listrightindent   \listrightindent = 0pt

				%

%-------------------------------------------------------------------------
						
						%time display in Spanish

						\def\mes{%
   			\ifcase\month
      \or Enero\or Febrero\or Marzo\or Abril\or Mayo\or Junio%
      \or Julio\or Agosto\or Septiembre\or Octubre\or Noviembre\or Diciembre%
   			\fi
							}%
					\def\timestring{\begingroup
   		\count0 = \time
   		\divide\count0 by 60
   		\count2 = \count0   % The hour, from zero to 23.
   		\count4 = \time
   		\multiply\count0 by 60
   		\advance\count4 by -\count0   % The minute, from zero to 59.
   		\ifnum\count4<10
     \toks1 = {0}%
   		\else
     \toks1 = {}%
   		\fi
   		\ifnum\count2<12
      \toks0 = {a.m.}%
   		\else
      \toks0 = {p.m.}%
      \advance\count2 by -12
   		\fi
   		\ifnum\count2=0
      \count2 = 12
   		\fi
   		\number\count2:\the\toks1 \number\count4 \thinspace \the\toks0
					\endgroup}%

%fecha, en castellano; fecha y hora,id.:

				%
				\newskip\abovelistskip      \abovelistskip = .5\baselineskip 
				\newskip\interitemskip      \interitemskip = 0pt
				\newskip\belowlistskip      \belowlistskip = .5\baselineskip
				\newdimen\listleftindent    \listleftindent = 0pt
				\newdimen\listrightindent   \listrightindent = 0pt

				%
%---------------------------------------------------------------
      \def\petit{\def\rm{\fam0\ninerm}%
      \textfont0=\ninerm \scriptfont0=\sixrm \scriptscriptfont0=\fiverm
       \textfont1=\ninei \scriptfont1=\sixi \scriptscriptfont1=\fivei
       \textfont2=\ninesy \scriptfont2=\sixsy \scriptscriptfont2=\fivesy
       \def\it{\fam\itfam\nineit}%
       \textfont\itfam=\nineit
       \def\sl{\fam\slfam\ninesl}%
       \textfont\slfam=\ninesl
       \def\bf{\fam\bffam\ninebf}%
       \textfont\bffam=\ninebf \scriptfont\bffam=\sixbf
       \scriptscriptfont\bffam=\fivebf
       \def\sans{\fam\sansfam\ninesans}%
       \textfont\sansfam=\ninesans \scriptfont\sansfam=\sixsans
       \scriptscriptfont\sansfam=\fivesans
       \def\tt{\fam\ttfam\ninett}%
       \textfont\ttfam=\ninett
       \normalbaselineskip=11pt
       \setbox\strutbox=\hbox{\vrule height7pt depth2pt width0pt}%
       \normalbaselines\rm
      \def\vec##1{{\textfont1=\tbms\scriptfont1=\tbmss
      \textfont0=\ninebf\scriptfont0=\sixbf
      \mathchoice{\hbox{$\displaystyle##1$}}{\hbox{$\textstyle##1$}}
      {\hbox{$\scriptstyle##1$}}{\hbox{$\scriptscriptstyle##1$}}}}}
%----------------------------------------------------------------------

%-------------------------------------------------------------------------
% footnotes macros:					

      \def\footnoterule{\kern-3pt\hrule width 2cm\kern2.6pt}
      \newdimen\oldparindent\oldparindent=1.5em
      \parindent=1.5em
 
      \newcount\footcount \footcount=0
      \def\advftncnt{\advance\footcount by1\global\footcount=\footcount}
      % automatically numbered footnotes, in petit
      \def\fnote#1{\advftncnt$^{\the\footcount}$\begingroup\petit
      \parfillskip=0pt plus 1fil
      \def\textindent##1{\hangindent0.5\oldparindent\noindent\hbox
      to0.5\oldparindent{##1\hss}\ignorespaces}%
      \vfootnote{$^{\the\footcount}$}{#1\nullbox{0mm}{2mm}{0mm}\vskip-9.69pt}\endgroup}
 %-------------------------------------------------------------------
%---------------------------------------------------------

%sets the text in a narrower box
        
 %------------------------------------------------------------------- 				

      \def\item#1{\par\noindent
      \hangindent6.5 mm\hangafter=0
      \llap{#1\enspace}\ignorespaces}
%-------------------------------------------------------------------
      
%note the difference with the TeXtbook \item, \itemitem, p. 355
%--------------------------------------------------------------------
      \def\leaderfill{\kern0.5em\leaders\hbox to 0.5em{\hss.\hss}\hfill\kern
      0.5em}
%-----------------------------------------------------------------------
						\def\hb{\hfill\break}
%-----------------------------------------------------------------------
						
						\def\tdots{$\,\ldots\,$}
%dots in text
    \def\centerrule#1{\centerline{\kern#1\hrulefill\kern#1}}
%a rule centered, with margins equal to #1

%--------------------------------------------------------------------------
%--------------------------------------------------------------------------
%boxing it:

      \def\boxit#1{\vbox{\hrule\hbox{\vrule\kern3pt
						\vbox{\kern3pt#1\kern3pt}\kern3pt\vrule}\hrule}}
      %puts a box around it

      \def\tightboxit#1{\vbox{\hrule\hbox{\vrule
						\vbox{#1}\vrule}\hrule}}
						%puts a tight box around it

      \def\looseboxit#1{\vbox{\hrule\hbox{\vrule\kern5pt
						\vbox{\kern5pt#1\kern5pt}\kern5pt\vrule}\hrule}}
      %puts a loose box around it

      \def\youboxit#1#2{\vbox{\hrule\hbox{\vrule\kern#2
						\vbox{\kern#2#1\kern#2}\kern#2\vrule}\hrule}}
      %puts a  box around #1 with margins specified by #2

%--------------------------------------------------------------------------

%various boxes, and tiles:

			\def\whitetile#1#2#3{\setbox0=\null
			\ht0=#1 \dp0=#2\wd0=#3 \setbox1=\hbox{\tightboxit{\box0}}\lower#2\box1}

%\nulltile is identical to \nullbox as described in 
% the \TeX book, p. 82 (cf. also p.312):
			\def\nullbox#1#2#3{\setbox0=\null
			\ht0=#1 \dp0=#2\wd0=#3\box0}

%------------------------------------------------------------------------------
%-----------------------------------------------------------------------------

   \def\permil{\leavevmode\hbox{\raise1ex%
   \hbox{$\scriptscriptstyle0$}\kern-0.2em%
   \raise0.4ex\hbox{\rm\char"2F}\kern-0.2em\hbox{$\scriptscriptstyle00$}}}
			%permil (one part in a thousand)

%------------------------------------------------------------------------------
\def\euro{$\tens{C}\kern-.65em\raise0.16em\hbox{{$\scriptstyle=$}}$}
%------------------------------------------------------------------------------

%-------------------------------------------------------------------------------

%common abreviations

\def\equ{\leavevmode Eq.}

\def\sect{\leavevmode Sect.}

%numbered:
\def\equn#1{\ifmmode \eqno{\rm #1}\else \equ~#1\fi}

%-------------------------------------------------------------
%en espa–ol:
\def\ecu{\leavevmode Ec.}

%numeradas:
\def\ecun#1{\ifmmode \eqno{\rm #1}\else \ecu~#1\fi}

%-------------------------------------------------------------------

\def\tev{\ifmmode \mathop{\rm TeV}\nolimits\else {\rm TeV}\fi}
\def\gev{\ifmmode \mathop{\rm GeV}\nolimits\else {\rm GeV}\fi}
\def\mev{\ifmmode \mathop{\rm MeV}\nolimits\else {\rm MeV}\fi}
\def\kev{\ifmmode \mathop{\rm keV}\nolimits\else {\rm keV}\fi}
\def\ev{\ifmmode \mathop{\rm eV}\nolimits\else {\rm eV}\fi}
\def\ryd{\ifmmode \mathop{\rm Ry}\nolimits\else {\rm Ry}\fi}
\def\angst{\ifmmode\mathop{\rm\AA}\nolimits\else {\rm \AA}\fi}

\def\degreec{\ifmmode\mathop{^\circ \rm C}\nolimits\else{$^\circ{\rm C}\;$}\fi}
\def\degreek{\ifmmode\mathop{^\circ \rm K}\nolimits\else{$^\circ{\rm K}\;$}\fi}
\def\degreef{\ifmmode\mathop{^\circ \rm F}\nolimits\else{$^\circ{\rm F}\;$}\fi}

\def\chidof{\ifmmode\mathop\chi^2/{\rm d.o.f.}\else $\chi^2/{\rm d.o.f.}\null$\fi}

\def\msbar{\ifmmode\mathop{\overline{\rm MS}}\else$\overline{\rm MS}$\null\fi}

\def\cmass{\ifmmode\mathop{\rm c.m.}\nolimits\else {\sl c.m.}\fi}
\def\lab{\ifmmode{\rm lab}\else {\sl lab.}\fi}

\def\degrees{\ifmmode{^\circ\,}\else $^\circ$\fi}
\def\feet{\ifmmode{\hbox{'}\,}\else '\fi}
\def\inches{\ifmmode{\hbox{"}\,}\else "\fi}

%all these may be used in math or text modes.

%-------------------------------------------------------------------------------

\def\TeX{T\kern-.1667em\lower.5ex\hbox{E}\kern-.125emX\null}%
\def\LaTeX{L\kern -.26em \raise .6ex \hbox{\sevenrm A}\kern -.15em \TeX}%
\def\AMSTeX{$\cal A\kern -.1667em
   \lower .5ex\hbox{$\cal M$}%
   \kern -.125em S$-\TeX
}%
\def\BibTeX{{\rm B\kern-.05em{\sevenrm I\kern-.025em B}\kern-.08em
    T\kern-.1667em\lower.7ex\hbox{E}\kern-.125emX}}%
\def\physmatex{P\kern-.14em\lower.5ex\hbox{\sevenrm H}ys
\kern -.35em \raise .6ex \hbox{{\sevenrm M}a}\kern -.15em
 T\kern-.1667em\lower.5ex\hbox{E}\kern-.125emX\null}%

%----------------------------------------------------------------------------
\def\ref#1{$^{[#1]}$\relax}
%references, in square brackets
%------------------------------------------------------------------------------   
%-------------------------------------------------------------------------------
%Journals

%-----------------------------------------------------------------------------
%---------------------------------------------------------------------------
% MATHEMATICAL MACROS

%(vector) arrow over character.

%allows stacking of 2 over 1.

%this allows to write under or above, the equality sign,
% with an underscore (_) order.

%the same with the "simeq" sign.

%the same with the "congruent" sign.

%the same under or above an arrow.

%the same with an identity sign 

%the same under, or above a similarity sign.

%the same under, or above an approx sign.

%------------------------------------------------------------------------
\def\dddoverdots{.\kern-.8pt .\kern-.8pt .}

%\dot#1, \ddot#1, \dddot#1 puts one, two, three dots over the character.
%\dot#1, \ddot#1 are standard TeX feature. \dddot#1 isnew.

\def\underdot#1{\mathord{\vtop to0pt{\ialign{##\crcr
$\hfil\displaystyle{#1}\hfil$\crcr\noalign{\kern1.5pt\nointerlineskip}
$\hfil\dot{}\kern1.5pt\hfil$\crcr}\vss}}}

\def\underddot#1{\mathord{\vtop to0pt{\ialign{##\crcr
$\hfil\displaystyle{#1}\hfil$\crcr\noalign{\kern1.5pt\nointerlineskip}
$\hfil{.\kern-0.7pt.}{}\kern1.5pt\hfil$\crcr}\vss}}}

\def\underdddot#1{\mathord{\vtop to0pt{\ialign{##\crcr
$\hfil\displaystyle{#1}\hfil$\crcr\noalign{\kern1.5pt\nointerlineskip}
$\hfil{.\kern-0.7pt .\kern-0.7pt .}{}\kern1.5pt\hfil$\crcr}\vss}}}

%this puts a dot, two dots, three dots under the character

\def\undertilde#1{\mathord{\vtop to0pt{\ialign{##\crcr
$\hfil\displaystyle{#1}\hfil$\crcr\noalign{\kern1.5pt\nointerlineskip}
$\hfil\tilde{}\kern1.5pt\hfil$\crcr}\vss}}}

%puts a tilde under the character

%this slashes the character; \slash is used for l.c. letters, \Slash for u.c ones
%---------------------------------------------
\def\lrvecd#1{\buildrel{\scriptstyle\leftrightarrow}\over{#1}}
\def\lrvecs#1{\buildrel{\scriptscriptstyle\leftrightarrow}\over{#1}}
\def\lrvecss#1{\buildrel{\scriptscriptstyle\leftrightarrow}\over{#1}}

\def\lrvec#1{\mathchoice{\lrvecd#1}{\lrvecd#1}{\lrvecs#1}{\lrvecss#1}}

%leftrightvector over character
%--------------------------------------

%leftvector over character
%----------------------------------------------------------------------
\def\ddal{\mathop{\vrule height 7pt depth0.2pt
\hbox{\vrule height 0.5pt depth0.2pt width 6.2pt}\vrule height 7pt depth0.2pt width0.8pt
\kern-7.4pt\hbox{\vrule height 7pt depth-6.7pt width 7.pt}}}
\def\sdal{\mathop{\kern0.1pt\vrule height 4.9pt depth0.15pt
\hbox{\vrule height 0.3pt depth0.15pt width 4.6pt}\vrule height 4.9pt depth0.15pt width0.7pt
\kern-5.7pt\hbox{\vrule height 4.9pt depth-4.7pt width 5.3pt}}}
\def\ssdal{\mathop{\kern0.1pt\vrule height 3.8pt depth0.1pt width0.2pt
\hbox{\vrule height 0.3pt depth0.1pt width 3.6pt}\vrule height 3.8pt depth0.1pt width0.5pt
\kern-4.4pt\hbox{\vrule height 4pt depth-3.9pt width 4.2pt}}}

\def\dal{\mathchoice{\ddal}{\ddal}{\sdal}{\ssdal}}

%this produces the d'Alembertian operator,
% with correct display, script and scriptscript sizes

%--------------------------------------
\def\dlambdab{\lambda\kern-1.3mm\hbox{\vrule width1mm height 1.5mm depth -1.4mm}}
\def\slambdab{\lambda\kern-1.16mm\hbox{\vrule width0.8mm height 1.1mm depth -1.mm}}
\def\sslambdab{\lambda\kern-1.04mm\hbox{\vrule width0.6mm height 0.82mm depth -0.7mm}}

%lambda bar (rationalized wavelength)
%------------------------------------

\mathchardef\lap='0001
%this produces the Laplacian (upright uppercase Delta)

%-------------------------------------------

%\carre produces a square, no shadowing but WITH scaling

\def\lsim{\mathop{\setbox0=\hbox{$\displaystyle 
\raise2.2pt\hbox{$\;<$}\kern-7.7pt\lower2.6pt\hbox{$\sim$}\;$}
\box0}}
\def\gsim{\mathop{\setbox0=\hbox{$\displaystyle 
\raise2.2pt\hbox{$\;>$}\kern-7.7pt\lower2.6pt\hbox{$\sim$}\;$}
\box0}}
%these two represent, respectively, "less than, or sim", and "biger than, or sim". 
%to write under these signs, use the following commands, with #1 whatever goes under.

\def\gsimsub#1{\mathord{\vtop to0pt{\ialign{##\crcr
$\hfil{{\mathop{\setbox0=\hbox{$\displaystyle 
\raise2.2pt\hbox{$\;>$}\kern-7.7pt\lower2.6pt\hbox{$\sim$}\;$}
\box0}}}\hfil$\crcr\noalign{\kern1.5pt\nointerlineskip}
$\hfil\scriptstyle{#1}{}\kern1.5pt\hfil$\crcr}\vss}}}

\def\lsimsub#1{\mathord{\vtop to0pt{\ialign{##\crcr
$\hfil\displaystyle{\mathop{\setbox0=\hbox{$\displaystyle 
\raise2.2pt\hbox{$\;<$}\kern-7.7pt\lower2.6pt\hbox{$\sim$}\;$}
\box0}}
\def\gsim{\mathop{\setbox0=\hbox{$\displaystyle 
\raise2.2pt\hbox{$\;>$}\kern-7.7pt\lower2.6pt\hbox{$\sim$}\;$}
\box0}}\hfil$\crcr\noalign{\kern1.5pt\nointerlineskip}
$\hfil\scriptstyle{#1}{}\kern1.5pt\hfil$\crcr}\vss}}}
%---------------------------------------------------------------

\def\imag{\mathop{\rm Im}}
%the real/imaginary functions

\def\ii{{\rm i}}
\def\dd{{\rm d}}
%i, d for sqrt(-1) and differential

\def\ee{{\rm e}}

%the number e, and Euler's gamma

%residue and principal part

% \sumprime#1#2, \intprime#1#2, primed sums and integrals with automatic math scaling.

%pincipal determination of argument, angle and logarithm

%angle

%argument

%mod. Compare with TeX \pmod, \bmod (with brackets)

%trace

%Spur

%these give gradient, divergence and curl (rot)

%the Spanish sin, and Gringo rot (curl)

%---------------------------------------------------------------------
%-------------------------------------------------------------------------------

%square black box
%------------------------------------------------------------

\def\frac#1#2{{#1\over#2}}
\def\dfrac#1#2{{\displaystyle{#1\over#2}}}
\def\tfrac#1#2{{\textstyle{#1\over#2}}}
\def\ffrac#1#2{\leavevmode
   \kern.1em \raise .5ex \hbox{\the\scriptfont0 #1}%
   \kern-.1em $/$%
   \kern-.15em \lower .25ex \hbox{\the\scriptfont0 #2}%
}%
%various forms of fractions
%-------------------------------

%binomial expressions
%-------------------------------------------------------------------
%--------------------------------------------------------------------
\def\oversetbrace#1\to#2{\overbrace{#2}^{#1}}
\def\undersetbrace#1\to#2{\underbrace{#2}_{#1}}
%various under/over braces
%compare with \overbrace, \underbrace

%--------------------------------------------------------------------

%----------------------------------------------------------------------

\def\rightcorner#1#2{\vrule height-2.4pt width#2 depth2.9pt
\vrule height #1 depth2.8pt width.5pt\kern-.6mm}

\def\rightroof#1#2{\vrule height 9pt depth -8.5pt width#2
\vrule height 8.8pt depth#1 width.5pt
\kern-.6mm}

%to be used with \llap, e.g., \leftcorner{3mm}{4mm}\llap M
%Convenient values for 1 character, e.g. M or p:
%\leftcorner{3mm}{4mm};\rightcorner{3mm}{4mm}
%\leftroof{0.5mm}{4mm};\rightroof{0.5mm}{4mm}
%-------------------------------------------------------------------------
%--------------------------------------------------------------------

\def\lcorner{\kern0.8mm\vrule height 1mm \phantom{.}\kern-1mm\vrule height0.4pt width1mm}
\def\rcorner{\vrule height0.4pt width0.1mm\kern-1mm\phantom{.}\vrule height 1mm}

\def\tlcorner{\kern0.8mm\vrule height 0.53mm \phantom{.}\kern-1mm\vrule height0.4pt width1mm}
\def\trcorner{\vrule height0.4pt width0.1mm\kern-1mm\phantom{.}\vrule height 0.53mm}

\def\upcorchfill{
$\lcorner\leaders\vrule height0.4pt\hfill
\leaders\vrule height0.4pt\hfill\rcorner\kern0.8mm $}

\def\tupcorchfill{
$\tlcorner\leaders\vrule height0.4pt\hfill
\leaders\vrule height0.4pt\hfill\trcorner\kern0.8mm $}
%---------------------------------------------------

\def\dundercontraction#1{\mathop{\vtop{\ialign{##\crcr
$\hfil\kern-.2mm{\displaystyle #1}\hfil$\crcr
\noalign{\kern2pt\nointerlineskip}\upcorchfill\crcr\noalign{\kern3pt}}}}\!\,}

\def\tundercontraction#1{\mathop{\vtop{\ialign{##\crcr
$\hfil\kern-.2mm{\displaystyle #1}\hfil$\crcr
\noalign{\kern2pt\nointerlineskip}\tupcorchfill\crcr\noalign{\kern3pt}}}}\!\,}

\def\sundercontraction#1{\mathop{\vtop{\ialign{##\crcr
$\hfil\kern-.2mm{\scriptstyle #1}\hfil$\crcr
\noalign{\kern2pt\nointerlineskip}\tupcorchfill\crcr\noalign{\kern3pt}}}}\!\,}
		
\def\ssundercontraction#1{\mathop{\vtop{\ialign{##\crcr
$\hfil\kern-.2mm{\scriptscriptstyle #1}\hfil$\crcr
\noalign{\kern2pt\nointerlineskip}\tupcorchfill\crcr\noalign{\kern3pt}}}}\!\,}

%undercontraction: display, text, script, scriptscript
% under -contractions, between
%symbols enclosed in braces.
%---------------------------------------------

\def\dundercontractionlimits#1{\mathop{\vtop{\ialign{##\crcr
$\hfil\kern-.2mm\displaystyle{#1}\hfil$\crcr
\noalign{\kern3pt\nointerlineskip}
\upcorchfill\crcr\noalign{\kern3pt}}}}\limits}

%\dundercontractionlimits: allows writing under contraction
% with command _{}, in display mode
%----------------------

\def\yundercorch#1#2#3{\mathord{\vtop to0pt{\ialign{##\crcr
$\hfil\displaystyle{#3}\hfil$\crcr\noalign{\kern1.5pt\nointerlineskip}
$\hfil\ycorch{#1}{#2}{}\kern1.5pt\hfil$\crcr}\vss}}}

\def\ycorch#1#2{\vrule height #1 \vrule height 0.4pt width #2\vrule height #1}

\def\youundercontract#1#2{\;\yundercorch{#1}{#2}{\phantom{,}}\kern-4pt\kern-#2}

%\youundercontract: you choose height (#1) and length (#2) of contraction,
%counting from the symbol right after \youundercontract. 
%Useful for nested contractions

%---------------------------------------------------------------------------------
\def\tlroof{\vrule height0.1mm depth 0.55mm \hbox{\vrule height0.4pt width 1mm}}
\def\trroof{\hbox{\vrule height0.4pt width1.mm}\vrule height 0.1mm depth0.55mm}

\def\lroof{\vrule height0.1mm depth 1mm \hbox{\vrule height0.4pt width 1mm}}
\def\rroof{\hbox{\vrule height0.4pt width1.mm}\vrule height 0.1mm depth1mm}

\def\tovercorchfill{
$\tlroof\leaders\vrule height0.4pt depth0mm\hfill
\leaders\vrule height0.4pt depth 0mm\hfill\trroof\kern0.8mm $}

\def\overcorchfill{
$\lroof\leaders\vrule height0.4pt depth0mm\hfill
\leaders\vrule height0.4pt depth 0mm\hfill\rroof\kern0.8mm $}

\def\dovercontraction#1{\mathop{\vbox{\ialign{##\crcr\noalign{\kern3pt}
\kern.6mm\overcorchfill\crcr\noalign{\kern2pt\nointerlineskip}
$\hfil\kern-0.4mm{\displaystyle #1}\hfil$\crcr}}}\!\,}

\def\tovercontraction#1{\mathop{\vbox{\ialign{##\crcr\noalign{\kern3pt}
\kern.6mm\tovercorchfill\crcr\noalign{\kern2pt\nointerlineskip}
$\hfil\kern-0.4mm{\displaystyle #1}\hfil$\crcr}}}\!\,}

\def\sovercontraction#1{\mathop{\vbox{\ialign{##\crcr\noalign{\kern3pt}
\kern.6mm\tovercorchfill\crcr\noalign{\kern2pt\nointerlineskip}
$\hfil\kern-0.4mm{\scriptstyle #1}\hfil$\crcr}}}\!\,}

\def\ssovercontraction#1{\mathop{\vbox{\ialign{##\crcr\noalign{\kern3pt}
\kern.6mm\tovercorchfill\crcr\noalign{\kern2pt\nointerlineskip}
$\hfil\kern-0.4mm{\scriptscriptstyle #1}\hfil$\crcr}}}\!\,}

%\overcontraction: display, text, script, scriptscript 
% contractions above the symbols enclosed in braces.
%------------------------------------

\def\dovercontractionlimits#1{\mathop{\vbox{\ialign{##\crcr\noalign{\kern3pt}
\overcorchfill\crcr\noalign{\kern3pt\nointerlineskip}
$\hfil\displaystyle{#1}\hfil$\crcr}}}\limits}

%\dovercontractionlimits: you may write over contraction with command ^{},
%in display mode
%-------------------

\def\yovercorch#1#2{\kern2pt\vrule height 0.1mm depth #1 \hbox{\vrule height 0.4pt width #2}
\kern-1.1mm\vrule height0.1mm depth #1}

\def\yabovecorch#1#2#3{\mathop{\vbox{\ialign{##\crcr\noalign{\kern3pt}
\yovercorch{#1}{#2}\crcr\noalign{\kern3pt\nointerlineskip}
$\hfil\displaystyle{#3}\hfil$\crcr}}}\limits}

\def\youovercontract#1#2{\yabovecorch{#1}{#2}{\phantom{I}}\kern-6pt\kern-#2}

%\youovercontract: you choose height (#1) and length (#2) of overcontraction,
%counting from the symbol right after \youovercontract. 
%Useful for nested contractions

%----------------------------------------------------------------------

%----------------------------------------------------------------------
\def\yoverc#1#2#3{\kern2pt\vrule height 2.6pt depth #1
 \hbox{\vrule height 2.6pt depth-2.1pt width #2}
\kern-4.2mm$\rightarrow#3$}
\def\yabovec#1#2#3#4{\mathop{\vbox{\ialign{##\crcr\noalign{\kern3pt}
\yoverc{#1}{#2}{#3}\crcr\noalign{\kern3pt\nointerlineskip}
$\displaystyle{#4}\hfil$\crcr}}}\limits}

%------------------------------------------------

\def\yc#1#2#3{\vrule height #1depth-2.4pt 
\vrule height 2.7pt depth-2.3pt width #2\kern-2.9mm \rightarrow #3}
\def\yunderc#1#2#3#4{\mathord{\vtop to0pt{\ialign{##\crcr
$\hfil\displaystyle{#4}\hfil$\crcr\noalign{\kern1.5pt\nointerlineskip}
$\hfil\yc{#1}{#2}{#3}\hfil$\crcr}\vss}}}

\def\underdeviatea#1#2#3#4{\setbox0=\hbox{$#3$}\;
\,\yunderc{#1}{#2}{\box0}{\kern -\wd0 #4}\nullbox{0mm}{#1}{0mm}}
%------------------------------------------------

%------------------------------------------------
%\underdeviatea, \overdeviatea : deviation, with arrows, of height #1, length#2,
%and with text along the arrow in #3, and text following in the old line in #4.

%------------------------------------------------
%----------------------------------------------------------------------
\def\yovercna#1#2#3{\kern2pt\vrule height 2.6pt depth #1
 \hbox{\vrule height 2.6pt depth-2.1pt width #2}
$#3$}
\def\yabovecna#1#2#3#4{\mathop{\vbox{\ialign{##\crcr\noalign{\kern3pt}
\yovercna{#1}{#2}{#3}\crcr\noalign{\kern3pt\nointerlineskip}
$\displaystyle{#4}\hfil$\crcr}}}\limits}

%------------------------------------------------------------------

\def\ycna#1#2#3{\vrule height #1depth-2.4pt
 \vrule height 2.75pt depth-2.25pt width #2 #3}
\def\yundercna#1#2#3#4{\mathord{\vtop to0pt{\ialign{##\crcr
$\hfil\displaystyle{#4}\hfil$\crcr\noalign{\kern0pt\nointerlineskip}
$\hfil\kern-1.2em\ycna{#1}{#2}{#3}\hfil$\crcr}\vss}}}

\def\underdeviate#1#2#3#4{
\setbox0=\hbox{$\;#3$}\kern1.4em\yundercna{#1}{#2}{\box0}{\kern -\wd0 #4}\nullbox{0mm}{#1}{0mm}}
%------------------------------------------------
%\underdeviate, \overdeviate: deviations without arrows

%----------------------------------------------------
%SOME CONVENIENT, SIMPLE FORMATS FOR BOOKS, BROCHURES AND PREPRINTS
%----------------------------------------------------

%------------------------------------
\def\nada{\phantom{M}\kern-1em}
\def\brochureendcover#1{\vfill\eject\pageno=1{\nada#1}\vfill\eject}

%write this after the cover page of a brochure, or preprint, for TWO SIDED printing
%omit for ONE SIDED PRINTING

%if you don't want to write anything on the back of the title page, leave it blank
%---------------------------------

%to end a brochure/paper with an even-numbered page.
%if using \brochureendcover: 

%if not using \endbrochurecover: write \bookendchapter

%------------------------------------------
%-----------------------------------------
%For book chapters, 
% numbering at bottom (leading page not numbered), and headlines:

\def\chapterb#1#2#3{\pageno#3
\headline={\ifodd\pageno{\ifnum\pageno=#3\hfil\else\rheadline\fi}
\else\lheadline\fi}
\def\rheadline{\hfil -{#2}-\hfil}
\def\lheadline{\hfil-{#1}-\hfil}
\footline={\hss{\rm -- \number\pageno\ --}\hss}
\voffset=2\baselineskip}

%numbering at top (leading page not numbered), and headlines:

\def\chaptert#1#2#3{\pageno#3
\headline={\ifodd\pageno{\ifnum\pageno=#3\hfil\else\trheadline\fi}
\else\tlheadline\fi}
\def\trheadline{\hfil -{#2}-\hfil\folio}
\def\tlheadline{\folio\hfil-{#1}-\hfil}
\nopagenumbers
\voffset=2\baselineskip}

%if you want the leading page to show a number:

\def\nchapterb#1#2#3{\pageno#3
\headline={\ifodd\pageno\rheadline
\else\lheadline\fi}
\def\rheadline{\hfil -{#2}-\hfil}
\def\lheadline{\hfil-{#1}-\hfil}
\footline={\hss{\rm -- \number\pageno\ --}\hss}
\voffset=2\baselineskip}

\def\nchaptert#1#2#3{\pageno#3
\headline={\ifodd\pageno{\ifnum\pageno=#3\hfil\folio\else\trheadline\fi}
\else\tlheadline\fi}
\def\trheadline{\hfil -{#2}-\hfil\folio}
\def\tlheadline{\folio\hfil-{#1}-\hfil}
\nopagenumbers
\voffset=2\baselineskip}

%-------------------------------------------------------
\def\bookendchapter{\ifodd\pageno\vfill\eject
\headline={\hfill}\footline={\hfill}\null \vfill\eject
 \else\vfill\eject \fi}
%at the end of the chapter, to finish with an even-numbered page.

\def\obookendchapter{\ifodd\pageno\vfill\eject
 \else\vfill\eject\null\headline={\hfill}\footline={\hfill} \vfill\eject\fi}
%at the end of the chapter, to finish with an odd-numbered page.

%-------------------------------------
%---------------------------------------
% For books:

%------------------------------------------------------------
\def\booksection#1{\setbox0=\vbox{\hsize=0.85\hsize\tolerance=500\raggedright\hfuzz=6mm
\noindent{\medfib #1}\medskip}\goodbreak\vskip0.6cm\box0
\nobreak
\noindent}
\def\booksubsection#1{\setbox0=\vbox{\hsize=0.85\hsize\tolerance=400\raggedright\hfuzz=4mm
\noindent{\fib #1}\smallskip}\goodbreak\vskip0.45cm\box0
\nobreak
\noindent}
%--------------------------------------------------

%--------------------------------------------------
% For brochures/papers:

%-------------------------------------------------
%For brochures/papers
% numbering at bottom (leading page not numbered), and headlines:

\def\brochureb#1#2#3{\pageno#3
\headline={\ifodd\pageno{\rheadline}
\else\lheadline\fi}
\def\rheadline{\hfil -{#2}-\hfil}
\def\lheadline{\hfil-{#1}-\hfil}
\footline={\hss -- \number\pageno\ --\hss}
\voffset=2\baselineskip}

%numbering at top (leading page not numbered), and headlines:

\def\brochuret#1#2#3{\pageno#3
\headline={\ifodd\pageno{\trheadline}
\else\tlheadline\fi}
\def\trheadline{\hfil -{#2}-\hfil\folio}
\def\tlheadline{\folio\hfil-{#1}-\hfil}
\nopagenumbers
\voffset=2\baselineskip}
%---------------------------
%brochure chapters, sections, subsections:

%--------------------------------------------

%--------------------------------------------
% Quotations:

 %#1 is text of quote; #2 is author of quote; #3 is reference for quote;
%#4 is size. The quote appears in \box0, to be positioned ad lib. 
%with quothr, author and reference are rightlined; with quothl, left-justified.

%---------------------------------------------------
% figure captions

%---------------------------------------------------

%captiontype

%-----------------------------------------------------

%abstracttype INCLUDES size of box (valid also for table of contents),  
%AND specification of \petit font.

\def\abstracttype#1{\hsize0.7\hsize\tolerance=800\hfuzz=0.5mm \noindent{\fib #1}\par
\medskip\petit}

%--------------------------------------------------

%------------------------------------------------------
\def\hb{\hfill\break}

%-----------------------------

%Macros for formatting text, addressed to mathematics 
%and physics books and papers. Some definitions are original; others have been 
%taken from eplain.tex (Karl Berry et al.), 
%manmac.tex, the \TeXtbook (D. Knuth), or Springer-Verlag macros.

%-----------------------------------------------------------------------------
%FONTS

 %for titles
%for titles
%-------------------------------
 %for reference in quotations; for running head
 %for text in quotations
\font\smallsc=cmcsc10 at 9pt %small caps
%----------------------------------------------------------------------
\font\fib=cmfib8
\font\medfib=cmfib8 at 9pt
\font\bigfib=cmfib8 at 12pt

%for sections and chapter headings
%--------------------------------------------------------------------
%Fonts for authors, in references

 %caps and small caps
%--------------------------------------------------------

%other fonts

\font\addressfont=cmbxti10 at 9pt%for (professional) address

 %text italic

%----------------------------------------------------------------
\catcode`@=11 % borrow the private macros of PLAIN (with care)

\newdimen\pagewidth \newdimen\pageheight \newdimen\ruleht
 \maxdepth=2.2pt  \parindent=3pc
\pagewidth=\hsize \pageheight=\vsize \ruleht=.4pt
\abovedisplayskip=6pt plus 3pt minus 1pt
\belowdisplayskip=6pt plus 3pt minus 1pt
\abovedisplayshortskip=0pt plus 3pt
\belowdisplayshortskip=4pt plus 3pt

\newinsert\margin
\dimen\margin=\maxdimen
%\count\margin=0 \skip\margin=0pt % marginal inserts take up no space

%----------------------------------------------------------

%margins:\topmargin=  ,\bottommargin=  ,\leftmargin=  ,\rightmargin=
%\advancetopmargin=  ,\advancebottommargin=  ,etc

%%Care should be exercised in putting these instructions
% AFTER any \magnification!

\newdimen\paperheight \paperheight = \vsize
\def\topmargin{\afterassignment\@finishtopmargin \dimen0}%
\def\@finishtopmargin{%
  \dimen2 = \voffset		% Remember the old \voffset.
  \voffset = \dimen0 \advance\voffset by -1in
  \advance\dimen2 by -\voffset	% Compute the change in \voffset.
  \advance\vsize by \dimen2	% Change type area accordingly.
}%
\def\advancetopmargin{%
  \dimen0 = 0pt \afterassignment\@finishadvancetopmargin \advance\dimen0
}%
\def\@finishadvancetopmargin{%
  \advance\voffset by \dimen0
  \advance\vsize by -\dimen0
}%
\def\bottommargin{\afterassignment\@finishbottommargin \dimen0}%
\def\@finishbottommargin{%
  \@computebottommargin		% Result in \dimen2.
  \advance\dimen2 by -\dimen0	% Compute the change in the bottom margin.
  \advance\vsize by \dimen2	% Change the type area.
}%
\def\advancebottommargin{%
  \dimen0 = 0pt \afterassignment\@finishadvancebottommargin \advance\dimen0
}%
\def\@finishadvancebottommargin{%
  \advance\vsize by -\dimen0
}%
\def\@computebottommargin{%
  \dimen2 = \paperheight	% The total paper size.
  \advance\dimen2 by -\vsize	% Less the text size.
  \advance\dimen2 by -\voffset	% Less the offset at the top.
  \advance\dimen2 by -1in	% Less the default offset.
}%
\newdimen\paperwidth \paperwidth = \hsize
\def\leftmargin{\afterassignment\@finishleftmargin \dimen0}%
\def\@finishleftmargin{%
  \dimen2 = \hoffset		% Remember the old \hoffset.
  \hoffset = \dimen0 \advance\hoffset by -1in
  \advance\dimen2 by -\hoffset	% Compute the change in \hoffset.
  \advance\hsize by \dimen2	% Change type area accordingly.
}%
\def\advanceleftmargin{%
  \dimen0 = 0pt \afterassignment\@finishadvanceleftmargin \advance\dimen0
}%
\def\@finishadvanceleftmargin{%
  \advance\hoffset by \dimen0
  \advance\hsize by -\dimen0
}%
\def\rightmargin{\afterassignment\@finishrightmargin \dimen0}%
\def\@finishrightmargin{%
  \@computerightmargin		% Result in \dimen2.
  \advance\dimen2 by -\dimen0	% Compute the change in the right margin.
  \advance\hsize by \dimen2	% Change the type area.
}%
\def\advancerightmargin{%
  \dimen0 = 0pt \afterassignment\@finishadvancerightmargin \advance\dimen0
}%
\def\@finishadvancerightmargin{%
  \advance\hsize by -\dimen0
}%
\def\@computerightmargin{%
  \dimen2 = \paperwidth		% The total paper size.
  \advance\dimen2 by -\hsize	% Less the text size.
  \advance\dimen2 by -\hoffset	% Less the offset at the left.
  \advance\dimen2 by -1in	% Less the default offset.
}%
%--------------------------------------------------------------------------

\def\onepageout#1{\shipout\vbox{ % here we define one page of output
    \offinterlineskip % butt the boxes together
    \vbox to 3pc{ % this part goes on top of the 44pc pages
      \iftitle % the next is used for title pages
        \global\titlefalse % reset the titlepage switch
        \setcornerrules % for camera alignment
      \else\ifodd\pageno \rightheadline\else\leftheadline\fi\fi
      \vfill} % this completes the \vbox to 3pc
    \vbox to \pageheight{
      \ifvoid\margin\else % marginal info is present
        \rlap{\kern31pc\vbox to\z@{\kern4pt\box\margin \vss}}\fi
      #1 % now insert the main information
      \ifvoid\footins\else % footnote info is present
        \vskip\skip\footins \kern-3pt
        \hrule height\ruleht width\pagewidth \kern-\ruleht \kern3pt
        \unvbox\footins\fi
      \boxmaxdepth=\maxdepth
      } % this completes the \vbox to \pageheight
    }
  \advancepageno}

\def\setcornerrules{\hbox to \pagewidth{\vrule width 1pc height\ruleht
    \hfil \vrule width 1pc}
  \hbox to \pagewidth{\llap{\sevenrm(page \folio)\kern1pc}%
    \vrule height1pc width\ruleht depth\z@
    \hfil \vrule width\ruleht depth\z@}}

\newbox\partialpage

%-----------------------------------------------------------------------
%----------------------------------------------------------------
%TeX units: 1 in=2.54 cm; 1 pc=4.218 mm; 1 pt= 0.351 mm
% dd=(1238/1157) pt; sp=(1/65536) pt; cc=12 dd.
%----------------------------------------------------------------
%-------------------------------------
% Page layout

%%Care should be exercised in putting the instructions
% below AFTER any \magnification!
%-----------------------------------

%for arbitrary dimensions:

%----------------------------------
%A4 size

%-------------------------------------
%-------------------------------------

\nopagenumbers   
%\rightline\timestamp
\rightline{FTUAM 06-07} 
\rightline{April 9,  2007}  
\rightline{arXiv: 0704.1201~~[hep-ph]}
%---------
\bigskip
\hrule height .3mm
\vskip.6cm
\centerline{{\bigfib Evaluation of the Axial  Vector Commutator Sum Rule}}
\bigskip 
\centerline{{\bigfib  for Pion-Pion
Scattering}}
\medskip
\centerrule{.7cm}
\vskip1cm
\setbox8=\vbox{\hsize65mm {\noindent\fib Stephen L. Adler}
\vskip .1cm
\noindent{\addressfont Institute for Advanced Study\hb
Einstein Drive\hb
Princeton, NJ  08540, USA}}
\centerline{\box8}
\smallskip
\setbox7=\vbox{\hsize65mm \fib and}
\centerline{\box7}
\smallskip
\setbox9=\vbox{\hsize65mm {\noindent\fib F. J.
Yndur\'ain}
\vskip .1cm
\noindent{\addressfont Departamento de F\'{\i}sica Te\'orica, C-XI\hb
 Universidad Aut\'onoma de Madrid,\hb
 Canto Blanco,\hb
E-28049, Madrid, Spain}\hb}
\smallskip
\centerline{\box9}
\bigskip

%--------
\setbox0=\vbox{\abstracttype{Abstract}
We consider the sum rule proposed by one of us (SLA),
obtained by taking the  expectation value of an axial vector commutator in
a state with one pion.
The sum rule relates the pion decay constant to integrals of pion-pion
 cross sections,
with one pion off the mass shell. We remark that recent data on pion-pion scattering allow
a precise evaluation of the sum rule. We also discuss
the related  Adler--Weisberger sum rule (obtained by
taking the expectation value of the same commutator in a state with one nucleon), especially in
connection with the  problem of extrapolation of the pion momentum off its mass shell. 
 We find, with current data, that both the pion-pion and 
pion-nucleon sum rules are satisfied to better than six percent, and 
we give detailed estimates of the experimental and extrapolation errors 
in the closure discrepancies.    
}
\centerline{\box0}
\brochureendcover{Typeset with \physmatex}
\brochureb{\smallsc s. l. adler and f. j.  yndur\'ain}{\smallsc
evaluation of the axial  vector commutator sum rule  for pion-pion
scattering}{1}

\booksection{1. Introduction.}

\noindent
In refs.~1,~2 a sum rule was obtained by taking matrix elements of the commutator,
$$[\chi_+(t),\chi_-(t)]=2I_3,
\equn{(1.1)}$$
between one-nucleon states. Here $\chi^\pm$ are the chiral charges,
$$\chi_\pm(x_0)=\int\dd^3{\bf x}\, A^{0}_\pm(x),
$$
 $I_3$ is the third component
of isospin and the $A^\mu_\pm$ are axial currents.
In terms of quarks,
$$A_+^\mu(x)=\bar{u}(x)\gamma^\mu\gamma_5d(x),\quad
A_-^\mu(x)=\bar{d}(x)\gamma^\mu\gamma_5u(x).$$

 One may note that, in QCD, the commutation
relation (1.1) is an exact theorem,  that follows from the global  symmetries of
the QCD Lagrangian; in fact, (1.1) may be obtained by integrating the
local axial vector commutation relation
$$\delta(x^0-y^0)[A_+^0(x),A_-^\nu(y)]=2\delta_4(x-y)V_3^\nu(x)
$$
with $V_3^\nu(x)=\tfrac{1}{2}(\bar{u}\gamma^\nu u-\bar{d}\gamma^\nu d)$ the third component of
the vector (isospin) current.
This commutation relation is exact in QCD, irrespective of the value of
the $u$, $d$ quark masses.

The sum rule obtained in refs.~1,~2
relates the  axial weak  charge, $g_A$, to integrals over pion-nucleon cross sections,
with the pion off-shell (and of zero mass),
$$1-\dfrac{1}{g_A^2}=\dfrac{2M_N^2}{g^2_rK_{NN\pi}(0)^2}
\,\dfrac{1}{\pi}\int_{(M_N+M_\pi)^2}^\infty\dfrac{\dd s}{s-M^2_N}\,
\left[\sigma^+_{0p}(s)-\sigma^-_{0p}(s)\right].
\equn{(1.2)}$$
Here, $\sigma^\pm_{0p}$ is the total cross section for scattering of a zero mass
$\pi^\pm$ on a proton, this latter on its mass shell (with averages over the spin of the nucleon
implicit).

Later, in ref.~3, it was noted that one can also take matrix
elements of (1.1) between
one pion states. In this way one obtains the relation
$$\dfrac{2}{g_A^2}=\dfrac{2M_N^2}{g^2_rK_{NN\pi}(0)^2}
\,\dfrac{1}{\pi}\int_{4M^2_\pi}^\infty\dfrac{\dd s}{s-M^2_\pi}\,
\left[\sigma^-_{0\pi}(s)-\sigma^+_{0\pi}(s)\right],
\equn{(1.3)}$$
where $\sigma^\pm_{0\pi}$ are the respective  total cross sections for scattering of a zero mass
$\pi^\pm$ on an on-shell $\pi^-$.
Both in (1.2) and (1.3) the {\sl exact} Goldberger--Treiman
relation\ref{4} (see Sect.~2 below for
the meaning of this)
$$1=\dfrac{M_Ng_A}{g_rK_{NN\pi}(0)F_\pi}
$$
was used to relate  the $\chi_\pm$ to the pion field.
In these formulas, $M_\pi=139.57\,\mev$ is the (charged) pion mass, $M_N$ the
nucleon mass,
$g_A$ is the weak axial coupling, $g_r$ the on-shell strong $NN\pi$ coupling,
and $K_{NN\pi}(0)$ the nucleon-nucleon-pion vertex, for a zero mass pion, normalized to
$K_{NN\pi}(M^2_\pi)=1$. $F_\pi$ is the pion decay constant.

One can rewrite (1.2), (1.3) in a form that is more convenient by
expressing the constant prefactor in terms of $F_\pi$, corresponding to
writing  the PCAC definition of the pion field in terms of $F_\pi$,
$$\partial\cdot A_\pm=\sqrt{2}\,F_\pi M^2_\pi\phi_\pm
\equn{(1.4)}$$
 ($\phi_\pm$ are
the  fields for $\pi^\pm$) and also expressing the sum rule in terms of
scattering amplitudes, rather than  cross sections. In this case we find
that (1.2), (1.3) are  replaced by the relations
$$g^2_A=1+8\pi F^2_\pi\int_{(M_N+M_\pi)^2}^\infty\dfrac{\dd s}{(s-M^2_N)^2}
\,\left\{\imag\, F_{\pi^+,p}(k^2=0;s,t=0)-\imag\, F_{\pi^-,p}(k^2=0;s,t=0)\right\}
\equn{(1.5)}$$
and
$$1=4\pi F^2_\pi\int_{4M^2_\pi}^\infty\dfrac{\dd s}{(s-M^2_\pi)^2}
\,\left\{\imag\, F_{\pi^+,\pi^-}(k^2=0;s,t=0)-\imag\, F_{\pi^-,\pi^-}(k^2=0;s,t=0)
\right\}.
\equn{(1.6)}$$
Here $F_{\pi^\pm,p}(k^2,s,t=0)$ stands for the forward pion-proton scattering
amplitude,
$$\pi^\pm(k) +N(p)\to\pi^\pm(k) +N(p)$$
with the nucleon $N(p)$ on its mass shell, $p^2=M^2_N$, while the momentum of the
pion is off-shell, $k^2=0$ in (1.5). Likewise, $F_{\pi^\pm,\pi^-}(k^2=0;s,t=0)$
is the forward scattering amplitude for a $\pi^\pm(k)$ with momentum
$k^2=0$ on a $\pi^-(p)$ on its mass shell, $p^2=M^2_\pi$:
$$\pi^\pm(k)+\pi^-(p)\to \pi^\pm(k)+\pi^-(p).$$

In the form (1.6), the pionic sum rule is seen as a remarkable
relation that allows one to calculate the pion decay constant in
terms of (off-shell) pion-pion scattering amplitudes. In ref.~3,
this sum rule for the pion case could not be fully evaluated. In
1965 data for $\pi\pi$ scattering in the S0 wave  were
so scanty that it was only possible to note that fulfillment of the
sum rule requires a strong $\pi\pi$ interaction in this S0 wave. The
situation has improved dramatically at present; precise and reliable
experimental data on the P and S0 waves for  $\pi\pi$ scattering,
and also on high energy scattering, have appeared in the last years,
which will allow us to present an evaluation of (1.3), (1.6) to a
few percent accuracy.

The nucleon sum rule,  known as the Adler--Weisberger sum rule,
 (1.2), (1.5) was already
evaluated in refs.~1,~2, 3; however,  we will present here a new calculation: the precision of
the pionic sum rule is such  that the question of the extrapolation to the off-shell
pion in (1.3), (1.6) becomes important,  and the corresponding discussion
is illuminated by  considering also
 what happens in the nucleon case.

In the present paper we will only consider the sum rules (1.5), (1.6); but,
by taking expectation values of (1.1) between kaon states one can get a sum
rule identical to (1.5), replacing the $g^2_A$ in the left hand side by zero,
the nucleon mass by the kaon mass, and pion-nucleon by
pion-kaon amplitudes: e.g., for a $K^+$ state, 
$$1=8\pi F^2_\pi\int_{(M_K+M_\pi)^2}^\infty\dfrac{\dd s}{(s-M^2_K)^2}
\,\left\{\imag\, F_{\pi^-,K^+}(k^2_\pi=0;s,t=0)-\imag\,
F_{\pi^+,K^+}(k^2_\pi=0;s,t=0)\right\}. \equn{(1.7)}$$

It is also possible to find a complementary relation to (1.7), 
but extrapolating in the mass of the kaon. 
Consider the currents $A_{K^+}^\mu=\bar{u}\gamma^\mu\gamma_5 s$, 
$A_{K^-}^\mu=\bar{s}\gamma^\mu\gamma_5 u$; 
 $A_{K^0}^\mu=\bar{d}\gamma^\mu\gamma_5 s$, 
$A_{\bar{K}^0}^\mu=\bar{s}\gamma^\mu\gamma_5 d$,
and the corresponding chiral charges, 
$\chi_{K^+}$, $\chi_{K^-}$;  $\chi_{K^0}$, $\chi_{\bar{K}^0}$. 
We have
$$[\chi_{K^+}, \chi_{K^-}]-[\chi_{K^0}, \chi_{\bar{K}^0}]=2I_3.$$
Taking expectation values 
on a $\pi^-$ state we now find the sum rule
$$\eqalign{
1=&\,4\pi F^2_K\int_{(M_K+M_\pi)^2}^\infty\dfrac{\dd s}{(s-M^2_\pi)^2}\,\big\{\imag\,
F_{K^+,\pi^-}(k_{K}^2=0;s,t=0)+F_{\bar{K}^0,\pi^-}(k_{K}^2=0;s,t=0)\cr
-&\,\imag\,F_{K^-,\pi^-}(k_K^2=0;s,t=0)-\imag\,F_{K^0,\pi^-}(k_K^2=0;s,t=0)\big\}\cr
=&\,8\pi F^2_K\int_{(M_K+M_\pi)^2}^\infty\dfrac{\dd s}{(s-M^2_\pi)^2}\,\big\{\imag\,
F_{K^+,\pi^-}(k_{K}^2=0;s,t=0)-
\imag\,F_{K^+,\pi^+}(k_K^2=0;s,t=0)\big\},\cr
}
 \equn{(1.8)}$$
the last relation using isospin and charge conjugation invariance.
We here leave the pion on its mass shell, and 
extrapolate in the kaon mass. 
Although there are some studies on pion-kaon scattering, variously 
using chiral perturbation theory, dispersive techniques and 
phenomenological information (see ref.~5 at low energies, and at
high energies, the 
Regge calculations of ref.~10),
the experimental information on $\pi K$ scattering amplitudes is
much less precise than that for $\pi\pi$ scattering.
However,  having two different
masses to extrapolate ($M_\pi$ and $M_K$) would
perhaps give some insight on the matter of extrapolation: the only differences
between (1.7) and (1.8)   are $F_K=113.00\pm1.06\,\mev$ instead of $F_\pi$, 
the different masses that are sent to zero, and the masses that appear in the denominators
in the integrals.

Similarly, as discussed by Weisberger, ref.~ 5,
 by  taking  appropriate chiral
charges in SU(3), and taking expectation values 
on nucleons one obtains relations involving kaon-nucleon
scattering. The pion-pion, pion-kaon or pion-nucleon sum rules require only an extrapolation in
the pion mass to yield an on-shell relation, while the kaon-nucleon [or kaon-pion, Eq.~(1.8)]
sum rule requires extrapolation in the much larger kaon mass to give
a physical relation, which is therefore expected to be less precise 
than the ones studied here, but could also give information 
on the question of extrapolation on the meson masses. 
Nevertheless, and as already stated we will, in the present paper, concern ourselves only with the 
pion-pion and pion-nucleon sum rules, for which good experimental data exist, and which 
require extrapolation  only to $M^2_\pi$; leaving the rest for future work.

\booksection{2. The Goldberger--Treiman relation}

\noindent Although the  Goldberger--Treiman relation\ref{4} is
not the object of this paper, we say a few words about it as we will
use some information connected with it. We first have
what may be called the exact or off-shell  Goldberger--Treiman relation,
$$\dfrac{M_Ng_A}{K_{NN\pi}(0)g_rF_\pi}=1.
\equn{(2.1)}$$
This involves the pion-nucleon coupling, with the pion off its mass shell
(with momentum $k^2=0$). One can  define the quantity
(the {\sl G.T. discrepancy})
$$\deltav_{\rm G.T.}\equiv1-\dfrac{M_Ng_A}{g_r F_\pi},
\equn{(2.2)}$$
which measures the validity of the  approximation $K_{NN\pi}(0)\simeq K_{NN\pi}(M^2_\pi)=1$
 that is used to get the on-shell Goldberger--Treiman
relation $F_\pi=M_N g_A/g_r$.

With the present values $g_A=1.2695\pm0.0029$, $F_\pi=92.42\pm0.26$
(both from the Particle Data Tables\ref{6}) and with
$M_N=938.9$ (average $n$-$p$ mass)
and taking
the on-shell pion nucleon coupling constant\ref{7}
$g_r=13.2\pm0.2$, the   relation (2.2) equals
$$\deltav_{\rm G.T.}=0.023\pm0.016,
\equn{(2.3)}$$
so the effect of approximating $K_{NN\pi}(0)$ by unity appears to be small.
 This corrects most of the
mismatch  studied by Pagels and Zepeda,\ref{8} which turns out to be
largely due to an underestimated $g_A$ (a possibility that they
actually considered); the remainder in (2.3) can easily be
attributed, as was done in ref.~8, to the contribution of the
$\pi(1300)$ resonance to $K_{NN\pi}(0)$, expected to be of
 $O[M^2_\pi/M^2_{\pi(1300)}]\sim1\%$.

One can look at the off-shell Goldberger--Treiman relation in a different way, as providing
the value of the
quantity
$K_{NN\pi}(0)$: (2.1)
 tells us that
$$K_{NN\pi}(0)=\dfrac{M_Ng_A}{F_\pi}=0.977\pm0.015. 
\equn{(2.4)}$$
In fact, from the careful analysis of Pagels and
Zepeda, it follows that most of the deviation of $K_{NN\pi}(0)$ from
unity is due to the fact that, since the pion is off-shell, and the
corresponding Green's function is not amputated, the $NN\pi$ vertex
must contain a factor $M^2_\pi\piv(0)$, where $\piv(k^2)$ is the
pion propagator normalized to
$1=(M^2_\pi-k^2)\piv(k^2)|_{k^2=M^2_\pi}$.
 Thus, one expects
$$K_{NN\pi}(0)\simeq M^2_\pi\piv(0),
\equn{(2.5)}$$ which will play a role in the extrapolation
discussion for the sum rules below. In fact,  the same factor
$M^2_\pi\piv(0)$ will appear in the sum rules because,  both in the
pion-pion and pion-nucleon cases, the propagator corresponding to
the off-shell pion line is not amputated.  Use of $\deltav_{\rm G.T.}$ 
to estimate off-shell extrapolation corrections has also been 
discussed by Dominguez.\ref{8}  

\booksection{3. Calculation of the sum rule on pions}
\vskip-0.5truecm
\booksubsection{3.1. The sum rules}

\noindent We  present here a sketch of the derivation of the sum rules,
for ease of reference; more details may be found in refs.~1,~2,~3.
We will treat in detail the pionic case derived in ref. 3, and indicate the modifications necessary
for the nucleon case.

We first take the expectation value of the commutator (1.1) between
physical $\pi^-$  states, and introduce a sum over a complete set of
states: we find
$$\eqalign{
\langle \pi^-(p')|2I_3 |\pi^-(p)\rangle=&\,-2\times2p_0\delta({\bf p}-{\bf p}')\cr
=&\,
\int\dd (q^2)\int\dfrac{\dd^3{\bf q}}{2q_0}
\sum_{\rm INT}\langle \pi^-(p')| \chi_+(t)|q;{\rm INT}\rangle
\langle q;{\rm INT} | \chi_-(t)| \pi^-(p)\rangle-(+\leftrightarrow -),\cr
}
\equn{(3.1)}
$$
where by $|q;{\rm INT}\rangle$ we denote a physical intermediate state
with total momentum $q$ and internal degrees of freedom ``INT".
Next, using the PCAC definition (1.4), one can relate
$$\langle q;{\rm INT} | \chi_\pm(x_0)| \pi^-(p)\rangle=
\dfrac{-\sqrt{2}\,\ii}{p_0-q_0}F_\pi M^2_\pi
\langle q;{\rm INT} |\int\dd^3 {\bf x}\,\phi_\pm(x)| \pi^-(p)\rangle
$$
with $\phi_\pm(x)$ the field operator for $\pi^\pm$.
Working in the infinite momentum frame (${\bf p}\to\infty$)
 gives the sum rule

$$1=2\pi F_\pi^2\int\dd s\,\dfrac{1}{(s-M^2_\pi)^2}
\sum_{\rm INT}\left\{|F(\pi^+(k^2=0),\pi^-(p)\to q;{\rm INT})|^2-
|F(\pi^-(k^2=0),\pi^-(p)\to q;{\rm INT})|^2\right\};
\equn{(3.2a)}$$
here $ s=q^2$,
and
 $$F(\pi^\pm(k^2=0),\pi^-(p)\to q;{\rm INT})=(2\pi)^{5/2}\langle q;{\rm INT} |
M^2_\pi\phi_\pm(0)| \pi^-(p)\rangle
\equn{(3.2b)}$$
is the amplitude for a pseudoscalar current,
with virtual four momentum $k$, $k^2=0$, to scatter off a physical $\pi^-$,
$p^2=M^2_\pi$, into the
physical intermediate state $|q;{\rm INT}\rangle$.
Using extended unitarity, this
 may  be written as the sum rule (1.6), 
which we repeat here in the form of a discrepancy, $\deltav_\pi=0$, with 
$$\deltav_\pi\equiv4\pi F^2_\pi\int_{4M^2_\pi}^\infty\dfrac{\dd s}{(s-M^2_\pi)^2}
\,\left\{\imag\, F_{\pi^+,\pi^-}(k^2=0;s,t=0)-\imag\, F_{\pi^-,\pi^-}(k^2=0;s,t=0)
\right\}-1.
\equn{(3.3)}$$  

In these formulas, the $F_{\pi^\pm,\pi^-}(k^2=0;s,t=0)$ are  the forward scattering amplitudes for
an off-shell pion with zero mass. For an {\sl on-shell} pion, the corresponding
scattering amplitude is obtained by replacing, in (3.2b),  
$$\eqalign{
F(\pi^\pm(k^2=0),\pi^-(p)\to q;{\rm INT})=&\,(2\pi)^{5/2}\langle q;{\rm INT} |
M^2_\pi\phi_\pm(0)| \pi^-(p)\rangle\cr
=
(2\pi)^{5/2}\langle q;{\rm INT} |
(M^2_\pi+\dal)\phi_\pm(0)| \pi^-(p)\rangle\Big|_{k^2=0}\to&\,
(2\pi)^{5/2}\langle q;{\rm INT} |
(M^2_\pi+\dal)\phi_\pm(0)| \pi^-(p)\rangle\Big|_{k^2=M^2_\pi}\cr
}
\equn{(3.4)}$$
and using unitarity to perform the sum over intermediate states.

In QCD one expects the mass scale for the internal dynamics to be given by a
parameter $\mu_0$ of the order of the
parameter $\lambdav\sim0.4\,\gev$ or the rho resonance mass, $M_\rho$; thus, to an error
 $M^2_\pi/\mu^2_0 \sim 10\%$  or smaller, we can relate (3.3) to
physical quantities  by approximating the off-shell scattering amplitudes
by the physical scattering amplitudes as in (3.4).

The sum rule (1.2), (1.5) on nucleons is derived in a similar manner. The only differences are
the
replacement of the  pion mass by the nucleon mass
 in the denominator corresponding to (3.3),
the different isospin of the proton (that results in a factor $-1/2$ with
respect to that for the
$\pi^-$) and that, due to the
existence of the neutron intermediate state, we find the extra term
proportional to $g_A^2$ in (1.5) because the proton-neutron matrix element of the
divergence
$\langle p|(\partial \cdot A|n\rangle$ is proportional to $g_A$.
The pion-kaon and kaon-pion sum rules are also similar to the pion-pion one. 

\booksubsection{3.2. The pion sum rule in the on-shell approximation}

\noindent
In the form (3.3), the sum rule is an {\sl exact} theorem, following from
the commutation relation (1.1) and the definition (1.4); but, of course,
direct comparison with experiment is precluded by the fact that
the amplitudes  that appear in (3.3) involve a pion off its mass shell.
As stated in the previous subsection, a first approximation is obtained
by neglecting the fact that (3.3) is defined for off-shell pseudoscalar currents,
i.e., working in the approximation (3.4). The sum rule is then,

$$1\simeq4\pi F^2_\pi\int_{4M^2_\pi}^\infty\dfrac{\dd s}{(s-M^2_\pi)^2}
\,\left\{\imag F_{\pi^-}(s)-\imag\, F_{\pi^+}(s)\right\}
\equn{(3.5)}$$
with $F_\pm(s)$  physical, forward $\pi^\pm\pi^-$ scattering amplitudes.
One can  write these scattering amplitudes in
terms of amplitudes with well-defined
isospin in the $t$ or $s$ channels,
$$F_{\pi^-}(s)-F_{\pi^+}(s)=F^{(I_t=1)}=
\tfrac{1}{3}F^{(I_s=0)}+\tfrac{1}{2}F^{(I_s=1)}-\tfrac{5}{6}F^{(I_s=0)};
\equn{(3.6)}$$
 the $F$ are normalized so that, for pions on
 their mass shell, and in the elastic region,
one has
$$F^{(I_s)}=2\dfrac{2s^{1/2}}{\pi k}\sum_l (2l+1)\sin\delta_l^{(I_s)}
\ee^{\ii \delta_l^{(I_s)}}.
\equn{(3.7)}$$
The factor 2 in front of the right hand side is due to identity of the particles and
$k=\tfrac{1}{2}\sqrt{s-4M^2_\pi}$ is the center of mass  momentum, for physical pions.

 In this case, we can use the precise
determinations of the pion-pion scattering amplitudes,
 obtained fitting  experimental data,
 that have been found recently\ref{9,10}
thanks to the availability of  very precise data: on the low energy S0 wave from kaon
decays  and, for the P wave, from determinations of the pion form factor.\ref{11} One finds,
for the contributions of the various waves to the right hand side of
(3.5) for
energy below 1420~MeV,\ref{9}
$$\matrix{
{\rm S0};\;s^{1/2}\leq 932\;\mev:&\quad 0.408\pm0.013\cr {\rm
S0};\;932\leq s^{1/2}\leq1420 \;\mev:&\quad 0.043\cr {\rm D0}:&\quad
0.097\pm0.003\cr {\rm P}:&\quad 0.403\pm0.003\cr {\rm F}:&\quad
0.0016\cr {\rm S2}:&\quad -0.090\pm0.005\cr {\rm D2}:&\quad
-0.0023;\cr} \equn{(3.8a)}$$ errors are only given for the more
significant pieces. The results are similar to those already
obtained in ref.~3 (although, of course, now much more precise) for
the contributions of P, D0, S2 waves. What is new is the
contribution of the S0 wave, which turns out to be the most
important of all, thus confirming the prediction in ref.~3 that an
important S0 wave contribution is needed to saturate the sum rule.

For the (Regge) contribution above $1420\,\mev$ we 
use the Regge formula  
$$\imag F^{(1)}(s)=(1.22\pm0.14)(s/1\;{\rm GeV}^2)^{0.42}
$$
and then find\ref{10}
$$s^{1/2}>1420\;\mev:\quad 0.167\pm0.017.
\equn{(3.8b)}$$
Altogether, the right hand side of (3.5) now reads
$$1.027\pm0.022,\quad{\rm i.e.},\quad\deltav_\pi=0.027\pm0.022.   
\equn{(3.9)}$$
The  error is due to the experimental
errors in the pion-pion scattering amplitudes in (3.8).
Therefore, we only have a discrepancy of
 $(2.7\pm2.2)\%$ 
 in the fulfillment
of the sum rule in this approximation.

A few words may be said on the smallness of the error in (3.9). This
is due to the fact that, as stated, recent experimental results
 have allowed us to get
very precise fits to data at low energy.
Moreover, the larger contributions to the final result
come from {\sl independent} sources, so one can add their errors quadratically.
And, finally, all the large contributions (S0, P, D0 waves and
Regge region) are positive: only the relatively small
S2 and D2 contributions produce  cancellations.
As we will see, the situation is less favourable for the sum rule
on nucleons, where large cancellations take place.

\break  
\booksubsection{3.3. Extrapolation}

\noindent
To improve the evaluation using (3.5) one can
 calculate by taking the recipe of ref.~3 for extrapolation, which
 takes into account threshold {\sl kinematic} effects
by  replacing (3.7), in the elastic region,\fnote{Actually, we make
the corresponding replacement up to the Regge region,
 $s^{1/2}\simeq1.42\,\gev$.}
   by
$$F_0^{(I_s)}=\dfrac{4s^{1/2}}{\pi k}\sum_l (2l+1)
\left[\dfrac{k^{(0)}}{k}\right]^{2l}\sin\delta_l^{(I_s)}
\ee^{\ii \delta_l^{(I_s)}},
\equn{(3.10)}$$
where $k^{(0)}=(s-M^2_\pi)/2s^{1/2}$ is the center of mass
 momentum for a   pion of zero mass incident on a target pion
that is on mass shell.
This does not take into account the effects of the extrapolation at high energies,
$s^{1/2}>1.42\,\gev$, likely below the 1\% level, which we neglect.
We have verified that this recipe works in a model calculation in which
the interactions are generated by effective Lagrangians
$$g_\rho(\vec{\phi}\times {\lrvec \partial}_\mu \vec{\phi})\rho_\mu,
\quad
g_f\vec{\phi}(\dal g_{\mu\nu}-\partial_\mu\partial_\nu)\vec{\phi}f_2^{\mu\nu},\;\dots
$$ coupling pions to
various resonances [$\rho$, $f_2(1275)$,\tdots]. In these models,
the off shell correction is valid not only at threshold but
throughout the resonance region, which justifies our using it here
in the elastic region.

The recipe in (3.10) then gives a discrepancy
$$\deltav_\pi=0.069\pm0.023.
\equn{(3.11)}$$
This {\sl deteriorates} the sum rule, which seems to imply that the dynamical
effects of the
extrapolation are not negligible. In fact, this could be expected; the
replacement (3.10) does not affect the S waves, in particular the S0 wave, which is
the one that contributes most to the sum rule.

A possible way to estimate at least part of the dynamical correction
is to assume it to be universal, and take it from the
Goldberger--Treiman relation: thus multiplying the right hand side
of (3.11) by the factor
 $K^2_{NN\pi}(0)=0.955\pm0.03$, as was done in ref. 3; see also ref. 8.   
 A motivation
for this was already given when we discussed the Goldberger--Treiman
relation (end of \sect~2);
 the motivation for this in ref.~3 was
the observation that in a field theory of pions and nucleons, a zero
mass pion must couple to a physical pion through a virtual nucleon
loop, and so the factor $K_{NN\pi}(0)$ should be present (of course,
both motivations are not exclusive, but complementary). In terms of
QCD, an analogous observation is that an off-shell pion couples to
both nucleons and pions through a coupling to a single quark line,
suggesting that a universal off-shell factor may be present. It
would be of interest to pursue this idea further within a QCD framework;
see also the discussion in \sect~5 below.

Including this  factor $K^2_{NN\pi}(0)$ improves the fulfillment of
the sum rule to
$$\deltav_\pi=0.021\pm0.023,  
\equn{(3.12)}$$
 an agreement as good as one can wish. However,  as
we will see in the case of the nucleon sum rule, following the same
procedure of including a $K^2_{NN\pi}(0)$ factor depreciates, rather
than improves, agreement with experiment.   Hence  a more
conservative procedure is to take the difference between (3.11) and
(3.12) as a measure of the uncertainty in the extrapolation
procedure and thus write for the discrepancy, Eq.~(3.3b),
$$\deltav_\pi=0.021\pm0.023\;\hbox{(Exp.)}\pm0.048\;\hbox{(Extr.)},  
\equn{(3.13)}$$
showing explicitly  the error arising from  experimental errors in the
pion-pion amplitudes and the estimated error of the
extrapolation procedure. This will be
 our final result for the sum rule on pions.

\booksubsection{3.4. Connection with  chiral perturbation theory}

\noindent Consider the forward dispersion relation for the
(physical) amplitude $F^{(I_t=1)}(s)$,
$${F}^{(I_t=1)}(s)=\dfrac{2s-4M^2_\pi}{\pi}\,\int_{M^2_\pi}^\infty\dd s'\,
\dfrac{\imag {F}^{(I_t=1)}(s')}{(s'-s)(s'+s-4M^2_\pi)}.
$$
Evaluating it at threshold, and writing $F^{(I_t=1)}(4M^2_\pi)$
in terms of  scattering lengths,
we find the so-called (first) Olsson sum rule\ref{12} 
[which is thus  an exact consequence of the 
dispersion relation, independent of the current algebra
commutator (1.1)]:  
$$2a_0^{(0)}-5a_0^{(2)}= 3M_\pi\int_{4M_\pi^2}^\infty \dd s\,
\dfrac{\imag F^{(I_t=1)}(s)}{s(s-4M_\pi^2)}. 
\equn{(3.14)}$$
 One can
evaluate the scattering lengths to lowest order (l.o.) in chiral
perturbation theory, i.e., to lowest order in $M^2_\pi$
(actually, this is strictly equivalent
 to the old Weinberg\ref{13}
evaluation) finding
$$2a_0^{(0)}-5a_0^{(2)}\big|_{\rm l.o.}=\dfrac{3M_\pi}{4\pi F^2_\pi}
$$
and, substituting in (3.14), the l.o. relation
$$1= 4\pi F^2_\pi\int_{4M_\pi^2}^\infty \dd s\,
\dfrac{\imag F^{(I_t=1)}(s)}{s(s-4M_\pi^2)}\Bigg|_{\rm l.o.}.
\equn{(3.15)}$$
To l.o. (which implies $M^2_\pi\approx0$) Eq.~(3.15)
is equivalent to (1.6)
 since
$$\int_{4M_\pi^2}^\infty \dd s\,
\dfrac{\imag F^{(I_t=1)}(s)}{s(s-4M_\pi^2)}\Bigg|_{\rm l.o.}\approx
\int_{4M_\pi^2}^\infty \dd s\,
\dfrac{\imag F^{(I_t=1)}(s)}{(s-M_\pi^2)^2}\Bigg|_{\rm l.o.}\approx
\int_{4M_\pi^2}^\infty \dd s\,
\dfrac{\imag F^{(I_t=1)}(k^2=0;s,t=0)}{(s-M_\pi^2)^2}\Bigg|_{\rm l.o.}.$$
The same conclusion is reached if we use the Froissart--Gribov representation
of the P wave scattering length.

Of course, (3.15) differs in status from (1.6) 
 as the latter is
valid exactly, whereas the l.o. expression
$2a_0^{(0)}-5a_0^{(2)}\big|_{\rm l.o.}=3M_\pi/4\pi F^2_\pi$ is known
to have large  loop corrections.\ref{14}

Another connection with results based on chiral perturbation theory
is obtained by remarking that, using chiral methods to two loops,
plus analyticity (in the form of Roy equations) and extra
experimental information,  has led Colangelo, Gasser and
Leutwyler\ref{15}
 to propose parametrizations of the
S0, S2 and P waves at low energy consistent with these requirements.
 Substituting them into Eq.~(3.5) gives the result
 $\deltav_\pi=0.046$, to be compared with what we
found using experimental data in Eq.~(3.9), $0.027\pm0.022$.   
Likewise,  implementing off-shell corrections as in (3.10) gives
$\deltav_\pi=0.086$ [to be now compared with $0.069\pm0.023$ from
Eq.~(3.11)] and, finally, applying the correction deduced from the
off-shell Goldberger--Treiman relation under the universality
assumption gives   $\deltav_\pi=0.033$, a number which is also
similar to what was found using experimental $\pi\pi$ data in
Eq.~(3.12), viz., $0.021\pm0.023$.\fnote{The numbers 0.046, 0.086   
and 0.033 were kindly communicated to us by H. Leutwyler\ref{16} as
an independent check on the 
 evaluation of \sect~3.2, using different
methods.}

We emphasize that it is not our intention in this paper to give 
a detailed review of either chiral perturbation theory or the Olsson 
sum rule, which are substantial topics  in their own right. 
We are only interested in elucidating the connection 
between the Olsson relation and the pion-pion sum rule (1.6), which, 
as far as we know, can only be established to leading order in 
chiral perturbation theory.  Tests of the Olsson relation by itself 
have been made extensively in the literature; see, for example, refs. 
~14,~15 for evaluations using two-loop chiral perturbation theory, and 
the papers of Kami\'nski, Pel\'aez and Yndur\'ain and of Pel\'aez and 
Yndur\'ain in ref.~9 for calculations using experimental data prior 
to 2006.  The Olsson relation has also been verified including more 
recent data (those of the NA48/2 collaboration, ref. ~11) as in our 
evaluation of the pion-pion sum rule here, by Kami\'nski, Pel\'aez, and 
Yndur\'ain (unpublished work in progress).

\booksection{4. The Adler--Weisberger sum rule}
\vskip-0.5truecm
\booksubsection{4.1. An approximate calculation}

\noindent
The  Adler--Weisberger sum rule  (1.2), (1.5),
has been evaluated in a number of papers (besides the original ones); a recent
article is
ref.~17.   If we  approximate the amplitudes in (1.5) by the scattering
amplitudes for a physical pion, $k^2=M^2_\pi$,
then we get the sum rule
$$g^2_A\simeq1+8\pi f^2_\pi\int_{(M_N+M_\pi)^2}^\infty\dfrac{\dd s}{(s-M^2_N)^2}
\,\Big\{\imag F_{\pi^+}(s)-\imag\, F_{\pi^-}(s)\Big\}
\equn{(4.1)}$$
with $F_{\pi^\pm}(s)$  
 the physical, forward $\pi^\pm p$ scattering amplitudes,
normalized so that the pion-proton cross sections are
$$\sigma_{\pi^\pm p}(s)=
\dfrac{4\pi^2}{\lambda^{1/2}(s,M^2_\pi,M^2_N)}\imag F_{\pi^\pm}(s),
$$
with the function $\lambda(a,b,c)$ as defined below.
In terms of $s$-channel isospin amplitudes, we have
$$F_{\pi^+}(s)- F_{\pi^-}(s)=\tfrac{2}{3}\Big\{F^{(I_s=3/2)}-F^{(I_s=1/2)}\Big\}
$$
where the partial wave expansion in the elastic region is
$$F^{(I)}(s)=\dfrac{2s^{1/2}}{\pi k}
\sum_l\left\{(l+1)\sin\delta^{(I)}_{l+}\ee^{\ii\delta^{(I)}_{l+}}
+l\sin\delta^{(I)}_{l-}\ee^{\ii\delta^{(I)}_{l-}} \right\};
\equn{(4.2)}$$
the phase shifts $\delta^{(I)}_{l\pm}$ correspond to
isospin $I$, orbital angular momentum $l$, and
total angular momentum
$j=l\pm\tfrac{1}{2}$. The center of mass momentum is 
$$k=\tfrac{1}{2}\sqrt{\dfrac{\lambda(s,M^2_\pi,M^2_N)}{s}},\quad
\lambda(a,b,c)=[a^2-(b+c)^2][a^2-(b-c)^2].
$$

The high energy part, $s^{1/2}>2374\;\mev$ 
 of the integral in (4.1)
is easily evaluated with the fit to $\pi N$ cross sections in
ref.~10:
 Regge formulas are good approximations for kinetic energies above 1~GeV.
We will use the Regge formula
$$\imag F_{\pi^+}(s)-\imag\, F_{\pi^-}(s)=(0.42\pm0.04)(s/1\;{\rm GeV}^2)^{0.42},
$$
and integrate this  from $s^{1/2}=2374\;\mev$ to infinity. 
This would give a result of $-0.091$. Alternatively, we can use numerical values of the cross 
sections given in ref.~18 from $s^{1/2}=2374\;\mev$ to  $s^{1/2}=3004\;\mev$, and use the Regge 
formula above this latter energy. This would give $-0.087$. We consider this last number to be the more 
reliable one and then write, taking errors into account, 
$$
8\pi f^2_\pi\int_{(2374\;\mev)^2}^{\infty}\dfrac{\dd s}{(s-M^2_N)^2}
\,\Big\{\imag F_{\pi^+}(s)-\imag\, F_{\pi^-}(s)\Big\}=-0.087\pm0.006.
\equn{(4.3a)}$$

 On the other hand, numerical evaluation of the low
energy piece, using the numerical cross sections collected in
ref.~18  gives
$$
8\pi f^2_\pi\int_{(M_N+M_\pi)^2}^{(2374\;\mev)^2}\dfrac{\dd
s}{(s-M^2_N)^2} \,\Big\{\imag F_{\pi^+}(s)-\imag\,
F_{\pi^-}(s)\Big\}=0.460\pm0.024. \equn{(4.3b)}$$ 
 This large error
is due to the fact that the number in the right hand side is the
difference between two large numbers: specifically,
$$0.46=1.71\;[{\rm from\;\pi^+}]-1.25\;[{\rm from\;\pi^-}].
\equn{(4.4)}$$

Substituting the value of $g_A$ the sum rule (4.1) reads
$$1.612\pm0.006=1+0.373\pm0.025.  
\equn{(4.5)}$$
We may define a discrepancy $\deltav_{\rm A.W.}$ as the difference
between $g^2_A$ and the right hand side of Eq.~(1.5),   
 and express the result in (4.5)
as a largish mismatch,
$$\deltav_{\rm A.W.}=0.239\pm0.025.$$

In the present case, there are various substantial cancellations:
as already stated, there are cancellations between
the $\pi^+p$ and $\pi^-p$ cross sections at low energy, but
 there is also a
cancellation between the low
energy region and higher energy
($s^{1/2}>1.390$~GeV) contributions.
For example, if we only integrated up to and including the $\Delta(3,3)$
resonance region, we would have obtained
$$
8\pi f^2_\pi\int_{(M_N+M_\pi)^2}^{(1390\;\mev)^2}\dfrac{\dd s}{(s-M^2_N)^2}
\,\Big\{\imag F_{\pi^+}(s)-\imag\, F_{\pi^-}(s)\Big\}=0.70. 
$$
These cancellations amplify the errors in the sum rule, and
indicate that the effects
  of the extrapolation to a zero mass pion, which
affect mostly  low energies, are now very
 important, as already remarked in refs.~1 and 3.

\booksubsection{4.2. Extrapolation}

\noindent
To perform the extrapolation, we  repeat the method used for
the case of the sum rule on pions, and replace
the expression (4.2) for the scattering amplitude by
$$F_0^{(I)}(s)=\dfrac{2s^{1/2}}{\pi k}
\sum_l\left[\dfrac{k^{(0)}}{k}\right]^{2l}
\left\{(l+1)\sin\delta^{(I)}_{l+}\ee^{\ii\delta^{(I)}_{l+}}
+l\sin\delta^{(I)}_{l-}\ee^{\ii\delta^{(I)}_{l-}} \right\}
\equn{(4.7)}$$ with $k^{(0)}$ the momentum for an incident pion of
zero mass,
 $$k^{(0)}=\tfrac{1}{2}\sqrt{\dfrac{\lambda(s,0,M^2_N)}{s}}.
$$
We integrate this  in the elastic region,
$s^{1/2}\lsim1.5\,\gev$, using the parametrizations of ref.~19.
These parametrizations have been obtained by fitting up to energies of,
respectively, 1.3~\gev\ and 1.38~\gev; however,
we have verified that they continue to
fit the experimental cross sections up to $\sim150~\mev$ above their nominal
maximum. Between these energies and 1.9~\gev, one can use a resonance saturation
model,  with the resonance parameters taken from the PDT;\ref{6} plus a
background, estimated as the tail of the lower energy resonances.
Above 1.9~\gev, there are not
well determined values for the
resonance parameters and, moreover, a resonance model
 will cease to be valid as one is entering the
Regge regime. Fortunately, one can likely neglect the
effects of the extrapolation above such energy, and so this will be  done here.

 The results one gets for the extrapolation correction are
$$\eqalign{0.225:\;\hbox{up to 1460~\mev};\quad -0.065:\;\hbox{resonances, from 1460~\mev};
\quad 0.005:\; \hbox{background, from 1460~\mev}.\cr}   
\equn{(4.8)}$$
Adding this and considering as an error estimate the variation of the result if we vary the
matching point of the parametrizations and the resonance model from 1460   
 to 1420~\mev\ or
1520~\mev,  we get
a correction of $0.165\pm0.009$, 
 and hence the sum rule becomes
$$1.612\pm0.006=1+(0.373\pm0.025)+(0.165\pm0.009)=1.538\pm0.034.  
\equn{(4.9)}$$
We have added the errors linearly, as they are clearly correlated.
The results show
 reasonable fulfillment of the Adler--Weisberger
sum rule, $\deltav_{\rm A.W.}=0.074\pm0.034$.

If we include a  global correction, as we did in the
pionic case,  multiplying the r.h.s. of (4.9) by the factor
 $K^2_{NN\pi}(0)=0.955\pm0.03$, the sum rule now deteriorates
to
$$1.612\pm0.006=1+[(0.373\pm0.025)+(0.165\pm0.009)]\times(0.955\pm0.030)
=1.514\pm0.038.
\equn{(4.10)}$$
 We can now  write our final result, as we did for the pionic case, as
(4.10), adding as an extra error the difference between (4.10) and
(4.9):
$$\deltav_{\rm A.W.}=0.098\pm0.006\;[g^2_A]\pm0.031\;\hbox{[Exp.]}\pm0.035\;\hbox{[Extr.]}
\equn{(4.11)}$$ 

To compare with the results obtained in refs. 1 and 3, we note that
these papers did not multiply through by a factor of $g_A^2=1.612$,
and  included the $K^2_{NN\pi}(0)$ extrapolation factor.  Thus, the
relevant number to use is  [cf. (4.10)]
 $(0.514 \pm 0.038)/1.612=0.319
\pm 0.024$. The comparable number in refs. 1 and 3 is
$(4M_N^2/g_r^2)(R_1+R_2+R_3) =0.254+0.155-0.061=0.348$, with a
roughly estimated error of $\pm 0.025$.  Hence our current
evaluation, and the 1965 evaluation of refs. 1 and 3, are in
satisfactory agreement.  This should come as no surprise, since good
pion-nucleon cross sections were already available in 1965.  What
has changed dramatically since then, and is the motivation for the
present  paper, is the status of data on pion-pion scattering.

\booksection{4.3. Connection with lowest order chiral perturbation theory}

\noindent As we did for the pion case, the Adler--Weisberger
relation can be connected with (lowest order) chiral perturbation
theory, by comparing e.g., (4.1) with the forward dispersion
relation for the $\pi p$ scattering amplitude for exchange of
isospin unity at threshold,\fnote{This is generally known  as the
Goldberger--Miyazawa--Oehme sum rule.\ref{20}}  evaluating the
amplitude at threshold in terms of the scattering lengths
combination $a_3-a_1$, and calculating the latter in terms of
$F_\pi$, as in the
  Tomozawa--Weinberg articles.\ref{21} Details of this  may be found in
ref.~17.

\booksection{5. Comments}

\noindent We have shown  that with current data, 
 the pion-pion sum rule, as well
as the pion-nucleon one, is satisfied to better than six percent.
To improve on this precision, it will be necessary to have an improved
understanding of dynamical extrapolation corrections that account
 for the appearance of a zero
mass, off-shell pion  in the sum rules.  We make two remarks in this
regard.  The first is that extrapolation of the incident pion to
zero mass is {\sl not} 
 the same as taking the chiral limit of QCD,
since the target pion or nucleon, and all intermediate state
particles, remain on mass shell.  The second is that while for a
generic pion interpolating field the results of this extrapolation
are not well-defined, the sum rules involve a very specific choice
of pion interpolating field: the divergence of the axial vector
current, which is a well defined entity in QCD, as are the on shell
pion-pion and pion-nucleon scattering amplitudes. Thus, the question
of estimating extrapolation corrections, that are needed for a very
accurate comparison of the sum rules with experiment, is a
well-posed one in QCD. Modern lattice 
methods may permit improvement on the
estimates that have been used here and in refs.~1 and 3.

In this respect, it is amusing to remark that, unlike the situation
in 1965, the precision of the pionic sum rule is now greater than
that of the pion-nucleon one. 
This is very likely due to the fact that the latter involves 
small differences of large numbers, so any
 small alteration is amplified here.

\vfill\eject
\booksection{Acknowledgements}

\noindent
We wish to thank H.~Leutwyler for supplying numbers that
give an independent check of the pion-pion sum rule evaluation, and
also for correspondence that assisted one of us (SLA) in
reconstructing the reasoning behind the extrapolation method used
for the pion-pion case in ref.~3. The work of SLA was supported in
part by the U.~S. Department of Energy under Grant
No.~DE-FG02-90ER40542.  FJY is grateful to J.~R.~Pel\'aez for
interesting discussions in the  preliminary stages of this paper,
and to  the Spanish DGI of the MEC under contract
FPA2003-04597 for financial support.

\booksection{References}
\item{1 }{Adler, S. L., {\sl Phys. Rev. Letters} {\bf 14}, 1051 (1965).}
\item{2 }{Weisberger, W. I., {\sl Phys. Rev. Letters} {\bf 14}, 1047 (1965).}
\item{3 }{Adler, S. L., {\sl Phys. Rev.} {\bf 140}, B736 (1965)
and (E) {\sl Phys. Rev.} {\bf149}, 1294 (1966) and
{\sl Phys. Rev.} {\bf175}, 2224 (1968).}
\item{4 }{Goldberger, M. L., and Treiman, S. B.,  {\sl Phys. Rev.}
 {\bf 109}, 193 (1958).}
\item{5 } References on pion-kaon scattering: 
 M. Jamin, J., Oller, J. A., and Pich, A., {\sl Nucl. Phys.} {\bf B587}, 331 (2000); G\'omez-Nicola, 
A., and Pel\'aez, J. R., {\sl Phys. Rev.} {\bf D65}, 054009 (2002); 
Descotes-Genon, S., and Moussallam, B., {\sl Eur. Phys. J.} {\bf C33},
409 (2004); Zhou, Z. Y., and  Zheng, H. Q., {\sl Nucl. Phys.} {\bf A775}, 
212 (2006).  Theoretical proposal of kaon-nucleon sum rules:  
{Weisberger, W. I., {\sl Phys. Rev.} {\bf 143}, 1302 (1966).}
\item{6 }{Particle Data Tables: Yao, W.-M., et al., {\sl Journal of Physics} {\bf G33}, 1
(2006).}
\item{7 }{Saino, M. E., hep-ph/9912337 and,
especially, Schindler, M. R., et al., nucl-th/0611083.} The second 
article reviews previous evaluations.   
\item{8 }{Pagels, H., and Zepeda, A.,  {\sl Phys. Rev.} {\bf D5}, 3262 (1972). 
For a more recent discussion of corrections to the Goldberger--Treiman 
relation, and their relation to off-shell extrapolation corrections, 
see Dominguez,~C.~A.,
 {\sl Phys. Rev. } {\bf D15}, 1350 (1977) and 
{\sl Phys. Rev.} {\bf D16}, 2320 (1977). }
\item{9 }{S0 wave, low energy ($s^{1/2}\leq932\,\mev$): Yndur\'ain, F. J.,
Garc\'{\i}a ~Mart\'{\i}n,~R., and
Pel\'aez, J. R., hep-ph/0701025.
P wave, low energy  ($s^{1/2}\leq932\,\mev$):
 de Troc\'oniz, J. F., and Yndur\'ain, F. J., {\sl Phys. Rev.},  {\bf D65},
093001,
 (2002),
and {\sl Phys. Rev.} {\bf D71}, 073008 (2005). S2 wave, D2 and F wave,
$s^{1/2}\leq1420\,\mev$:  Pel\'aez, J. R., and Yndur\'ain,~F.~J.,
{\sl Phys. Rev.} {\bf D71}, 074016 (2005). S0 and P waves, medium energy
(up to $s^{1/2}\leq1420\,\mev$);
 D0 wave:
Kami\'nski, R.,  Pel\'aez, J. R., and Yndur\'ain,~F.~J.,
 {\sl Phys.Rev.} {\bf D74}, 014001
(2006)  and
(E) {\bf D74}, 079903 (2006).}
\item{10}{High energy, $s^{1/2}>1420\,\mev$: Pel\'aez, J.~R., and
Yndur\'ain, F. J., {\sl Phys. Rev.} {\bf D69}, 114001 (2004).
Actually, in the present calculation one uses the slightly
 improved Regge parametrization in the
paper of
Kami\'nski, Pel\'aez and Yndur\'ain in  ref.~9;
 see also
Cudell, J. R., et al., {\sl Phys. Letters} 
{\bf B587}, 78 (2004) and Pel\'aez, J.~R., in {\sl Proc. Blois. Conf.
on  Elastic and Diffractive Scattering} (hep-ph/0510005)
 for the pion-nucleon case.}  
\item{11}{{\sl Kl4 decays}\/: Rosselet, L., et al., {\sl Phys. Rev.} {\bf D15}, 574  (1977);
Pislak, S.,  et al.,  {\sl
Phys. Rev. Lett.}, {\bf 87}, 221801 (2001);
NA48/2 ( CERN/SPS experiment);
Bloch-Devaux, B., presented at QCD06 in Montpellier
(France), 3-7 July 2006 and
Masetti, L.,  presented at ICHEP06 in Moscow (Russia), 26 July to 2 August 2006.
{\sl $K\to2\pi$ decays}\/: Aloisio, A., et al., {\sl Phys. Letters}, {\bf B538}, 21  (2002).
{\sl Pion form factor data}\/:
Novosibirsk, $\rho$ region: L.~M.~Barkov et al., {\sl Nucl. Phys.} {\bf B256}, 365 (1985);
  R.~R.~Akhmetshin et al.,
{\sl Phys. Letters} {\bf B527}, 161
(2002).  {\sl $\tau$ decays}\/:
ALEPH: R. Barate et al., {\sl Z. Phys.} {\bf C76}, 15 (1997); OPAL:
K.~Ackerstaff et al., {\sl Eur. Phys. J.} {\bf C7}, 571 (1999);
 CLEO: S.~Anderson et al., {\sl Phys. Rev.},
 {\bf D61}, 112002 (2000).}
\item{12}{Olsson, M. G., {\sl Phys. Rev.}{\bf 162}, 1338 (1967).}
\item{13}{Weinberg, S., {\sl Phys. Rev. Letters}  {\bf 17}, 616 (1966).}
\item{14}{Gasser, J., and Leutwyler, H., {\sl Ann. Phys.} (N.Y.) {\bf 158}, 14
 (1984).}
\item{15}{Colangelo, G., Gasser, J.,  and Leutwyler, H.,
 {\sl Nucl. Phys.} {\bf B603},  125, (2001). See also
Ananthanarayan, B., et al., {\sl Phys. Rep.}, {\bf 353}, 207,  (2001).
The Roy equations were derived in
Roy, S. M., {\sl Phys. Letters}, {\bf 36B}, 353,  (1971).}
\item{16} {Leutwyler, H. (private communication).}
\item{17}{Kondratyuk, S., et al., {\sl Nucl. Phys.} {\bf A736}, 339 (2004).}
\item{18}{The COMPAS Group compilations can be traced from Hagiwara, K., et al., {\sl
Phys. Rev.} {\bf D66}, 010001 (2002).}
\item{19}{Rowe, G., Salomon, M., and Landau, R. H., {\sl Phys. Rev.} {\bf C18},
584 (1978); Ebrahim, A. A., and Peterson,~R.~J., {\sl Phys. Rev.} {\bf C54}, 2499
(1996).}
\item{20}{Goldberger, M. L, Miyazawa, H., and Oehme, R., {\sl Phys. Rev.} {\bf 99}, 986 (1955).}  For a very recent evaluation and further 
references, see Abaev, V. V., Mets\"a, P., and Sainio, M. E., arXiv:0704.3167 [hep-ph].  
\item{21}{Tomozawa, Y., {\sl Nuovo Cimento} {\bf 46A}, 707 (1966); Weinberg, S., {\sl Phys. Rev.
Lett.} {\bf 19}, 616 (1966).}

\bye